\documentclass[twocolumn]{aastex63}
\usepackage{tablefootnote}
\usepackage{booktabs}        
\usepackage{amsmath}
\usepackage{makecell}
\usepackage{multirow}
\usepackage[utf8]{inputenc}
\usepackage[english]{babel}

\DeclareMathAlphabet{\pazocal}{OMS}{zplm}{m}{n}
\newcommand{\unif}{\pazocal{U}}
\makeatletter
\newcommand*{\linktocite}[2]{%
  \hyper@natlinkstart{#1}#2\hyper@natlinkend}
 \newcommand\addressresponse[1]{#1}
\makeatother
\accepted{December 9, 2021}
\submitjournal{The Astrophysical Journal}

\shorttitle{Confronting iron rain on WASP-76b}
\shortauthors{Savel et al.}
\graphicspath{{./}{figures/}}

\begin{document}

\title{No umbrella needed: Confronting the hypothesis of iron rain on WASP-76b with post-processed general circulation models}

\correspondingauthor{Arjun B. Savel}
\email{asavel@umd.edu}

\author[0000-0002-0786-7307]{Arjun B. Savel}
\affiliation{Department of Astronomy, University of Maryland, College Park, MD 20742, USA}

\author[0000-0002-1337-9051]{Eliza M.-R. Kempton}
\affiliation{Department of Astronomy, University of Maryland, College Park, MD 20742, USA}

\author[0000-0002-2110-6694]{Matej Malik}
\affiliation{Department of Astronomy, University of Maryland, College Park, MD 20742, USA}

\author[0000-0002-9258-5311]{Thaddeus D. Komacek}
\affiliation{Department of Astronomy, University of Maryland, College Park, MD 20742, USA}
\affiliation{Department of the Geophysical Sciences, The University of Chicago, Chicago, IL 60637, USA}

\author[0000-0003-4733-6532]{Jacob L. Bean}
\affiliation{Department of Astronomy \& Astrophysics, University of Chicago, Chicago, IL 60637, USA}

\author[0000-0002-2739-1465]{Erin M. May}
\affiliation{Johns Hopkins APL, 11100 Johns Hopkins Road, Laurel, MD 20723, USA}

\author[0000-0002-7352-7941]{Kevin B. Stevenson}
\affiliation{Johns Hopkins APL, 11100 Johns Hopkins Road, Laurel, MD 20723, USA}

\author[0000-0003-4241-7413]{Megan Mansfield}
\affiliation{Department of the Geophysical Sciences, The University of Chicago, Chicago, IL 60637, USA}

\author[0000-0003-3963-9672]{Emily Rauscher}
\affiliation{Department of Astronomy, University of Michigan, 1085 South University Avenue, Ann Arbor, MI 48109, USA}




\begin{abstract}

High-resolution spectra are unique indicators of three-dimensional processes in exoplanetary atmospheres. For instance, in 2020, Ehrenreich et al.\ reported transmission spectra from the ESPRESSO spectrograph yielding an anomalously large Doppler blueshift from the ultra-hot Jupiter WASP-76b. Interpretations of these observations invoke toy model depictions of gas-phase iron condensation in lower-temperature regions of the planet's atmosphere. In this work, we forward model the atmosphere of WASP-76b with double-gray general circulation models (GCMs) and ray-striking radiative transfer to diagnose the planet's high-resolution transmission spectrum. We confirm that a physical mechanism driving strong east-west asymmetries across the terminator must exist to reproduce large Doppler blueshifts in WASP-76b's transmission spectrum. We identify low atmospheric drag and a deep radiative-convective boundary as necessary components of our GCM to produce this asymmetry (the latter is consistent with existing Spitzer phase curves). However, we cannot reproduce either the magnitude or the time-dependence of the WASP-76b Doppler signature with gas-phase iron condensation alone. Instead, we find that high-altitude, optically thick clouds composed of $\rm Al_2O_3$, Fe, or $\rm Mg_2SiO_4$ provide reasonable fits to the Ehrenreich et al.\ observations --- with marginal contributions from condensation. This fit is further improved by allowing a small orbital eccentricity (\addressresponse{$e \approx 0.017$}), consistent with prior WASP-76b orbital constraints. We additionally validate our forward-modeled spectra by reproducing lines of nearly all species detected in WASP-76b by Tabernero et al. 2021. Our procedure's success in diagnosing phase-resolved Doppler shifts demonstrates the benefits of physical, self-consistent, three-dimensional simulations in modeling high-resolution spectra of exoplanet atmospheres.

\end{abstract}



\section{Introduction} \label{sec:intro}
Decades after the discovery of the first exoplanet \citep{mayor1995jupiter}, a harbinger of the population of ``hot Jupiters,'' observational, theoretical, and laboratory efforts have elucidated some of the key processes and features relevant to these planets \citep[e.g.,][]{showman2002atmospheric,cooper2006dynamics, knutson2007map,fortney2008unified,crossfield2010new, rauscher2012general, dobbs2013three,demory2013inference,kataria2015atmospheric, mendoncca2016thor, fleury2019photochemistry, winter2020stellar}. In the past five years, however, an even more extreme class of planet --- the ``ultra-hot Jupiter'' \citep{bell2018increased, arcangeli2018h, parmentier2018thermal} --- has emerged, once again requiring the construction of new theoretical frameworks. Featuring persistent dayside temperatures in excess of 2200~K, ultra-hot Jupiter atmospheres are thought to include thermal dissociation of molecules and thermal ionization of metals. The presence of gas-phase metals at low pressures within these atmospheres also renders ultra-hot Jupiters notably susceptible to temperature inversions by virtue of strong optical absorption of host star light \citep[e.g.,][]{lothringer2019influence}. In short, observations of ultra-hot Jupiters have revealed them to be sites for extreme physics, consequently requiring significant model updates to understand them in detail.

\begin{figure*}
    \centering
    \includegraphics[scale=0.25, trim=0 90 0 90, clip]{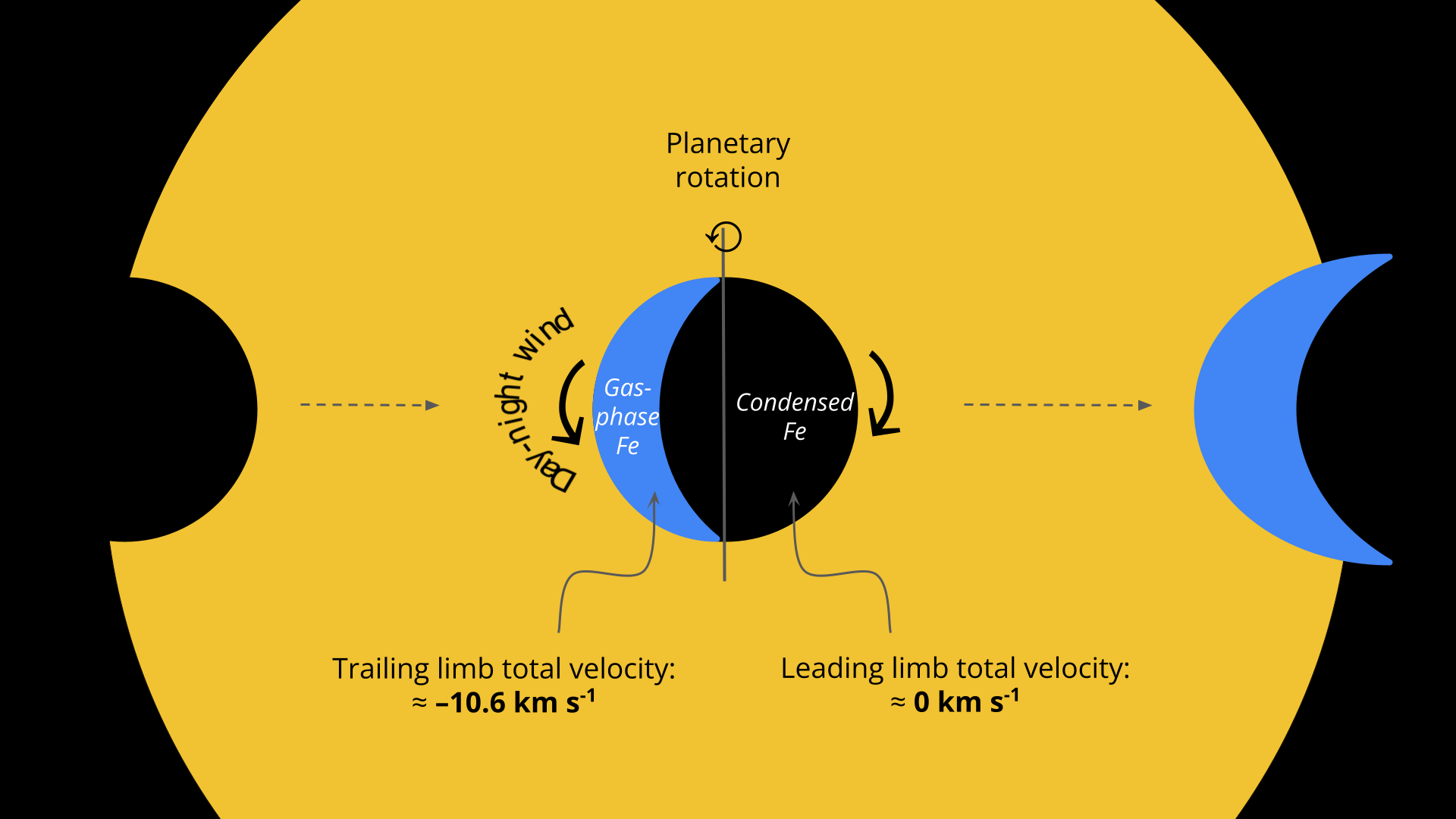}
    \caption{\addressresponse{A schematic diagram of the ``toy model'' introduced in \cite{ehrenreich2020nightside}. WASP-76b is pictured at three different phases (left to right: ingress, center of transit, and egress). Colder regions of WASP-76b's atmosphere, where Fe condenses and produces no absorption, are the primary regions probed at ingress. Over the course of its transit, the planet rotates and brings hotter regions, which contain gas-phase iron, into view for the observer. This geometry makes the strong combined velocity of the trailing limb ($\approx -5.3$~km\,s$^{-1}$ from winds, $\approx -5.3$~km\,s$^{-1}$ from rotation) observable as a Doppler blueshift by mid-transit. Atmosphere size not to scale.}}
    \label{fig:toy_model}
\end{figure*}

One particularly intriguing aspect of ultra-hot Jupiter atmospheres is their three-dimensional (3-D) structure. Such highly irradiated objects are predicted to have very strong day-night temperature gradients, and perhaps chemical gradients, too \citep[e.g.,][]{showman2009atmospheric, rauscher2012general, kataria2016wide}; these atmospheres' short chemical timescales relative to dynamical timescales should drive them to chemical equilibrium \citep{baeyens2021mixing}. These features make it challenging to justify treating such planets as 1-D objects, and the departures from 1-D behavior are especially apparent in certain current and future \citep{feng20202d,taylor2020understanding, plurial2020bias,macdonald2020cold} observations, including those at high spectral resolution, which are the subject of this paper.  

To account for the unique 3-D properties of ultra-hot Jupiters, various updates to ``traditional'' hot Jupiter general circulation models (GCMs) have been required. For instance, recent models have incorporated the cooling and heating effects of hydrogen dissociation and recombination \citep{tan2019atmospheric, mansfield2020evidence}. Moreover, the Lorentz forces and Ohmic dissipation related to free electrons in weakly ionized atmospheres have been captured to varying degrees of sophistication in GCM frameworks, often as a Rayleigh drag force \citep[e.g.,][]{perna2010magnetic,rauscher2013three, komacek2016atmospheric} or by coupling to the magnetic induction equation \citep{batygin2013magnetically,rogers2014magnetohydrodynamic,hindle2019shallow,hindle2021magnetic}. Finally, the net thermal flux at the lower boundary can substantially impact the GCM-predicted thermal structure and circulation higher in the atmosphere \citep[e.g.,][]{rauscher2012general, showman2015three}, making this parameter a useful tuning knob to explore the effect of differing interior entropies on global dynamics.

In terms of observing the unique characteristics of ultra-hot Jupiters, ground-based high-resolution spectroscopy offers a wealth of information related to their thermal, chemical, aerosol, and dynamical (i.e., wind) structures and properties \citep[e.g.,][]{snellen2013finding, kempton2014high, brogi2016rotation, hood2020prospects}. 
Like traditional transmission spectroscopy, observations at high spectral resolution make use of the fact that host star light filtered through an exoplanet's atmosphere will be preferentially absorbed by certain chemical species at certain wavelengths. Instead of necessarily dividing in-transit by out-of-transit spectra, high-resolution spectra can be extracted by noting that the stellar lines (and telluric lines from the Earth's atmosphere) will be essentially stationary during the planet's transit, modulo the host star's planet-induced acceleration relative to the Earth. In contrast, lines associated with the planet can be individually resolved and will Doppler shift over the course of the planet's transit because of the planet's orbital motion, winds, and rotation \citep{snellen2010orbital}. Ultra-hot Jupiters are therefore favorable targets for high-resolution spectroscopy; in addition to being likely to transit \citep[e.g.,][]{seagroves2003detection}, their short orbital periods ensure that their lines undergo appreciable Doppler shifts over the course of their transit, making the planetary spectrum easier to disentangle. Moreover, ultra-hot Jupiters have large atmospheric scale heights due to their low mean molecular weight and high temperatures, which increase their signal strengths in transmission.

Although hot Jupiter spectra derived from 1-D and 3-D modeling tend to generally agree in their interpretation of current low-resolution data \citep{fortney2010transmission}, significant departures --- containing information about wind fields, thermal profiles, and planetary rotation --- begin to crop up at higher resolution \citep{kempton2012constraining, showman2012doppler,kempton2014high, flowers2019high, beltz2020significant}. Together, then, GCMs and high-resolution radiative transfer modeling can work synergistically to generate spectra that consider the full 3-D nature of the underlying atmosphere.

A prime example of the unique benefits of the intersection of the 3-D nature of (ultra-)hot exoplanet atmospheres and high-resolution spectroscopy lies in the ultra-hot Jupiter WASP-76b, a gas giant orbiting an F7 star \citep{west2016three} that is well-studied at lower resolution \citep{fu2017statistical, tsiaras2018population,fisher2018retrieval,Edwards2018ares,vonessen2020stis, fu2020hubble}. Using the high-resolution (R~$\approx$~138,000) ESPRESSO spectrograph on the VLT \citep{pepe2010espresso, pepe2013espresso}, the \cite{ehrenreich2020nightside} team was able to produce novel, high-SNR, phase-resolved transmission spectra of this target across two separate transits. Curiously, an anomalous Doppler signature in the planet's transmission spectrum was detected: between 0 and $-5$~km\,s$^{-1}$ at ingress, but roughly $-11$~km\,s$^{-1}$ by egress. These speeds far exceed the few km\,s$^{-1}$ planet-frame velocities detected on other planets \citep{snellen2010orbital, brogi2016rotation}. The detection team attributes this variable and strong blueshift to an asymmetric distribution of atomic iron in the planet's atmosphere, with a considerable amount of iron existing in the gas phase east of the substellar point but cooling and condensing out as it makes its way to the much colder nightside. This asymmetry would cause a progressively blueshifted signal as the gas-phase iron region rotates into view over the course of the planet's transit. \cite{ehrenreich2020nightside} posit that their signal is composed of two independent Doppler components: solid-body rotation of $\pm$~5.3~km\,s$^{-1}$, the tidally locked equatorial velocity of the planet, and a uniform day-night wind contributing an additional $-5.3$~km\,s$^{-1}$ across both limbs, with gas phase iron only present on the evening limb of the planet \addressresponse{(Figure~\ref{fig:toy_model})}. \addressresponse{Hence, the approaching limb would exhibit a blueshift from both rotation and winds totaling --10.6~km\,s$^{-1}$, and the receding limb would produce no Doppler signal, as it would contain no gas-phase iron to absorb starlight.}

The \cite{ehrenreich2020nightside} ``toy model,'' as they put it, is supported by a separate, archival analysis of HARPS data \citep{kesseli2021confirmation}, which similarly finds increasing blueshift over the course of WASP-76b's transit. Additionally, the magnitude of the blueshift and speed of the day-night flow are confirmed by \cite{tabernero2021espresso} and \cite{seidel2021wasp}, who perform in-depth secondary analyses of the \cite{ehrenreich2020nightside} dataset. These works also produce rich datasets in their own right, with the \cite{tabernero2021espresso} study reporting the detection of multiple chemical species (Li I, Na I, Mg I, Ca II, Mn I, K I), their atmospheric heights, and their corresponding blueshifts.

The aforementioned iron chemistry gradient interpretation would be the first of its kind among exoplanet atmospheres. However, the necessity of a chemical gradient to explain the extant observations has been questioned by the forward models of \cite{wardenier2021wasp76}, who self-consistently calculate transmission spectra from a 3-D model of this planet and find that either iron condensation or a significant temperature asymmetry could match the \cite{ehrenreich2020nightside} data. Furthermore, Fe clouds are not necessarily favored in hot Jupiter atmospheres; microphysics models indicate that the nucleation rate of Fe is low, causing Fe clouds to be sequestered deep in the atmosphere \citep{gao2020aerosol, gao2021universal}.

In this paper, in addition to seeking to examine the iron rain hypothesis, we also aim to understand the physical, chemical, and dynamic processes at play in the atmosphere of WASP-76b. We do so with a suite of GCM simulations, applying a spectroscopic post-processing scheme to evaluate the likelihood of several plausible physical scenarios in the face of the constraints provided by the \cite{ehrenreich2020nightside}, \cite{tabernero2021espresso}, and \cite{kesseli2021confirmation} data. Our companion paper, \linktocite{maykomacek2021spitzer}{May \& Komacek et~al.} (\citeyear{maykomacek2021spitzer}), examines the dynamics of this planet with the aid of Spitzer phase curves; this paper focuses on the detected blueshift signature and chemical species by post-processing GCM simulations to produce high-resolution transmission spectra. Together, these works aim to compare the results of photometric phase curves and high-resolution transmission spectroscopy to paint a coherent picture of WASP-76b's 3-D atmosphere.

\begin{figure*}
\centering 
\includegraphics[scale=0.741, trim=0 0 0 0, clip]{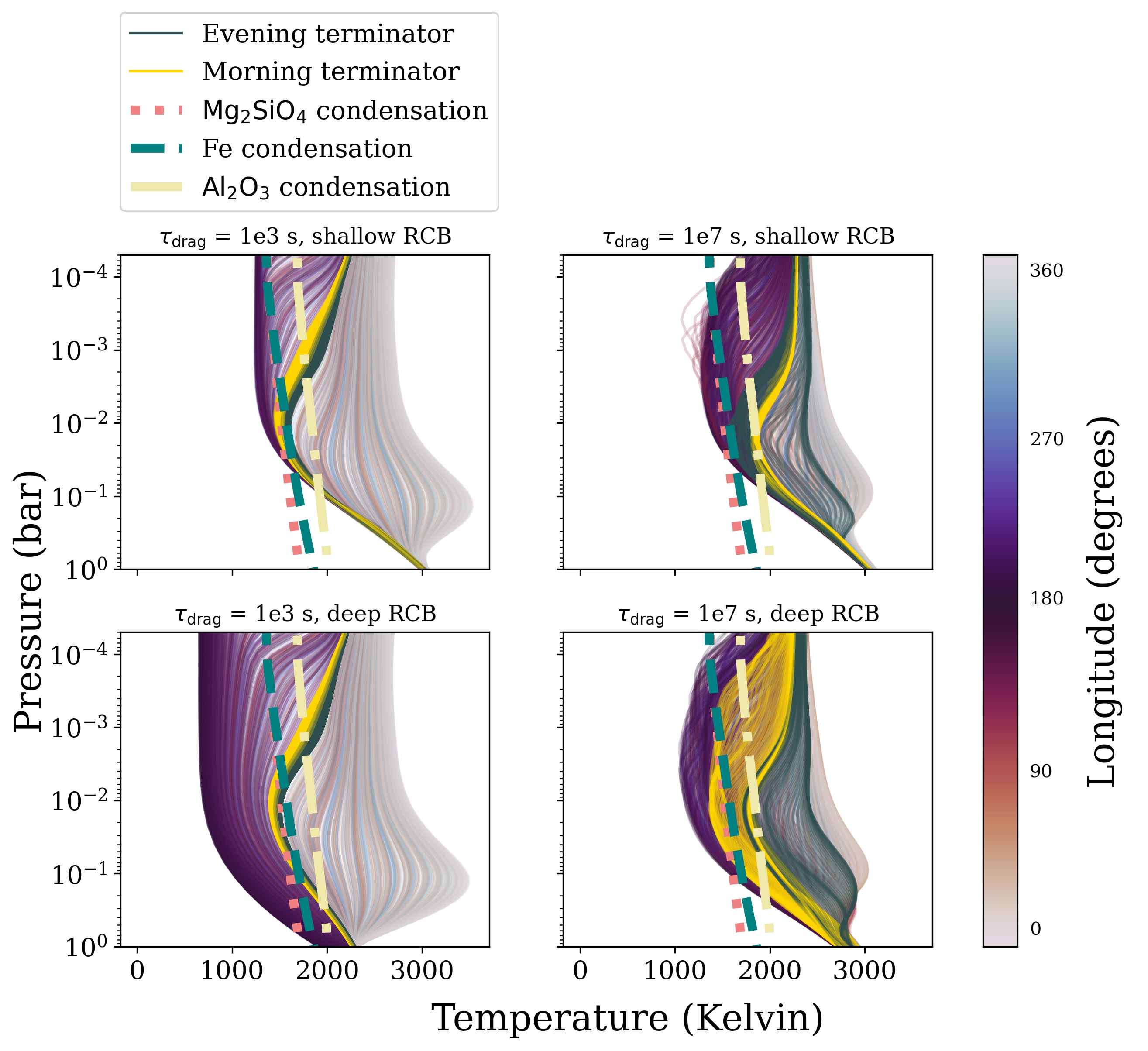}
\caption{1-D temperature-pressure (T-P) profiles of WASP-76b as output by our 3-D GCMs, with the condensation curves of $\rm Mg_2SiO_4$, Fe, and $\rm Al_2O_3$ from \cite{mbarek2016clouds} and terminators overplotted. Though the GCMs' outputs extend to 100 bars, our radiative transfer is limited to the (plotted) region at pressures less than 1 bar. Every 20th profile is plotted for ease of visualization, resulting in 902 samples of 18048 total latitude-longitude points. Regions near the substellar point (0$\degr$ / 360$\degr$) tend to be hotter on average than the rest of the planet, and regions near the antistellar point ($180\degr$) tend to be cooler on average than the rest of the planet. Significant scatter at a given longitude is controlled by latitudinal variation. The GCMs with deeper RCBs tend to have cooler nightsides than those with shallower RCBs, though all sets have regions in which Fe, $\rm Mg_2SiO_4$, and $\rm Al_2O_3$ condense (assuming equilibrium chemistry). Models with weaker drag (i.e., longer drag timescales) tend to produce much greater asymmetry between the east and west planetary limbs.}
\label{fig: t-p profs}
\end{figure*}

\begin{figure*}
\centering
\includegraphics[scale=0.86, trim=180 0 30 0, clip]{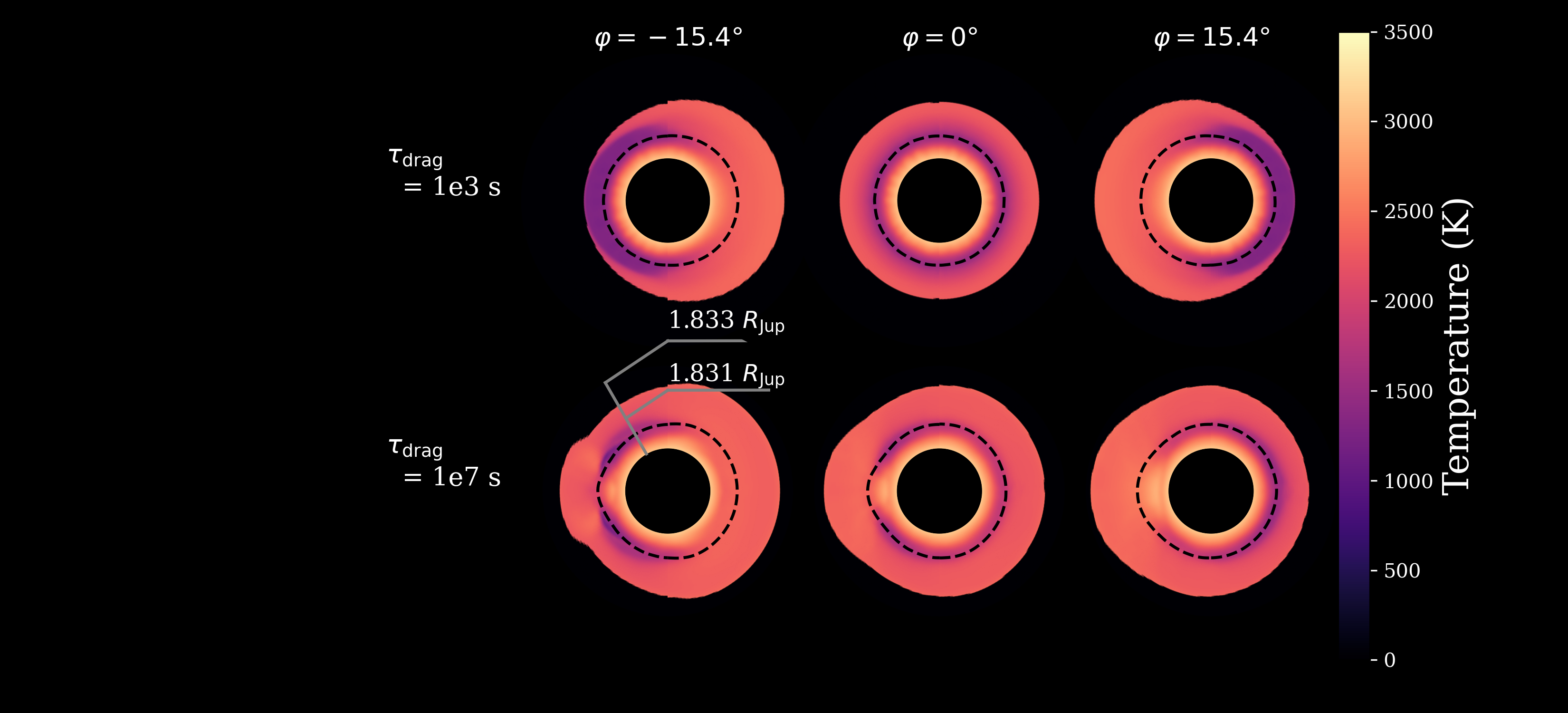}
\caption{Maps of atmospheric temperature at the terminator of WASP-76b for weak ($1\times 10^7$ s) and strong ($1\times 10^3$ s) drag timescales and three phases (ingress, center of transit, and egress). The planet core is not shown to scale, and we restrict our visualization to the region that is probed by high-resolution transmission spectroscopy ($\lesssim$ 1 bar --- our model domain slightly exceeds 1 bar in some regions, as our post-processed atmosphere is interpolated onto a grid evenly spaced in altitude). Black dashed lines correspond to the millibar isobar. As also shown in Figure \ref{fig:abund_maps}, the strong drag case enforces more overall east-west symmetry in the atmosphere of WASP-76b. The temperature inversion that can be seen in Figure \ref{fig: t-p profs} is also present here --- as would be expected for an ultra-hot Jupiter with high-altitude optical absorbers.}
\label{fig:temp_map}
\end{figure*}

This paper is organized as follows. Section~\ref{sec:model} describes the setup of our GCM and radiative transfer models, as well as the various physical scenarios we explore. Section~\ref{analysis} then lays out the analysis that we perform on the resultant spectra, which allows us to make direct comparisons to the \cite{ehrenreich2020nightside} and \cite{tabernero2021espresso} observational work. We present these comparisons in Section~\ref{results}, along with physical motivation for the individual effects included in our modeling results. Next, Section~\ref{discussion} briefly explores further alternative explanations for the anomalous blueshift and places our results in the context of other work. Finally, we summarize our conclusions in Section~\ref{conclusion}.

\begin{table}
\caption{Model parameters\footnote{Stellar and planetary parameters in the top six rows are taken from \cite{ehrenreich2020nightside}.}}
\label{table:parameters}
\centering
\begin{tabular}{ccc}
\toprule
\textbf{Parameter} & \textbf{Value(s)} &
\textbf{Unit}
    \\

\bottomrule
    $R_{\rm P}$ & 1.30$\times 10^8$ & m\\
    $g$ & 6.4 & $\rm m/s^2$\\
    $R_*$ & 1.22$\times 10^9$ & m\\
    $P_{\rm orb}$ & 1.81 & days \\
    $u_1$, $u_2$\footnote{Limb-darkening coefficients assume a quadratic law.} & 0.393, 0.219 & N/A \\
    $T_{\rm eff,*}$ & 6329 & K\\
   \\  \hline \\
    $\log_{10}\tau_{\rm drag}$ &  3, 4, 5, 6, 7 & s\\
    RCB & Shallow, deep & N/A\\
    Cloud composition\footnote{All clouds are modeled as fully optically thick.}  & Fe, $\rm Mg_2SiO_4$, $\rm Al_2O_3$ & N/A\\
    Max. cloud extent  & 1, 5, 10 & scale heights\\
    Condensation\footnote{Gas-phase iron condensation is treated independently from cloud formation.} & on, off  & N/A\\
    Orbital phase, $\varphi$ & ($-15.4$, 15.4) & degrees\\
    Orbital eccentricity, $e$ & (0.0, 0.05) & N/A\\
    Lon. of periastron, $\omega$ & (0, 360) & degrees
\\ \hline

\end{tabular}
\end{table}

\section{Model description}\label{sec:model}
\subsection{GCM}
As described above, we aim to produce physically motivated, self-consistent comparisons to the \cite{ehrenreich2020nightside} data by spectrally post-processing the outputs of a GCM. We use the same GCM model introduced and described in detail in our companion paper, which also includes the full GCM parameter assumptions and values (\linktocite{maykomacek2021spitzer}{May \& Komacek et~al.} \citeyear{maykomacek2021spitzer}, Table 3). Broadly, we use the MITgcm \citep{adcroft2004implementation} to solve the primitive equations of meteorology on a cubed-sphere grid with a horizontal resolution of C48 and 70 vertical layers, extending evenly in log-pressure from 100 bars to 10 microbars. To connect the modeled atmospheric circulation to heating and cooling within the atmosphere, the GCM is coupled with a two-stream, double-gray radiative transfer scheme with the TWOSTR \citep{kylling1995reliable} package of DISORT \citep{stamnes1988numerically}. A crucial update warranted for this project is the inclusion of local cooling where hydrogen dissociates, relevant to the intense dayside stellar irradiation of ultra-hot Jupiters, and local heating at locations where atomic hydrogen recombines, relevant to the much cooler limbs and nightsides of ultra-hot Jupiters \citep{tan2019atmospheric, mansfield2020evidence}.

As described in our companion paper (\linktocite{maykomacek2021spitzer}{May \& Komacek et~al.} \citeyear{maykomacek2021spitzer}), our GCMs (Figure \ref{fig: t-p profs}) incorporate a spatially constant Rayleigh drag force proportional to $\bf{v}_{\rm wind}/\tau_{\rm drag}$, for horizontal wind velocity $\bf{v}_{\rm wind}$ and drag timescale $\tau_{\rm drag}$. Our grid of GCMs encompasses a parameter sweep of drag timescales and two endmembers of radiative-convective boundary (RCB) depth, with the aim of exploring a wide range of possible dynamical states and interior states of WASP-76b. Our models are computed over a parameter sweep of drag timescales ($\tau_{\rm drag}$ $\in$ [$1 \times 10^3$s, $1 \times 10^7$s], with five values sampled evenly in log space), motivated by the unknown strength of turbulence and magnetic forces associated with the ionized dayside. The longest drag timescales correspond to atmospheres that have similar temperature structures and wind speeds to those with no applied frictional drag. We model upward heat fluxes from the deep planetary interior by including a ``surface'' beneath our model domain that does not interact with the atmospheric flow aside from heating it. The upward fluxes for our limiting cases of RCB depth are $3543.75~\mathrm{W}~\mathrm{m}^{-2} $ ($T_{\rm int}$= 500~K) and $4.474 \times 10^{7}~\mathrm{W}~\mathrm{m}^{-2}$ ($T_{\rm int}$= 5300~K) for the deep and shallow RCB cases, respectively.

\subsection{Radiative transfer model}
\subsubsection{Overview}\label{sec:radiative transfer overview}
For our study, we make use of a line-of-sight ray-striking radiative transfer model that includes Doppler effects, inherited from \cite{kempton2012constraining}, \citet{kempton2014high}, \cite{rauscher2014atmospheric}, and \cite{flowers2019high}. We begin with the GCM output, which we interpolate from a grid evenly spaced in $\log$ pressure onto a grid evenly spaced in altitude --- thus ensuring that our simulated rays propagate in straight lines. We further only consider the GCM output at pressures generally greater than 1 bar (the cut is made in altitude, not pressure, so the bottom-most pressure value is weakly latitude/longitude dependent), as the atmosphere is fully optically thick in transmission geometry at pressures greater than this value (Figure~\ref{fig:temp_map}). We convert between pressure and altitude assuming hydrostatic equilibrium, accounting for the non-uniform mean molecular weight between GCM grid cells due to the thermal dissociation of H$_2$ in hotter regions of the atmosphere. Our calculation assumes a nominal atmosphere in cooler regions that is 83.6\% H$_2$, 16\% He, and 0.4\% metal-bearing molecules (the latter having an average mean molecular weight $\mu = 12$) by number.

Accounting for the 3-D geometry of our GCM structures, we calculate the slant optical depth,
\begin{equation}
    \tau_{\lambda} = \int \kappa_{\lambda} ds,
\end{equation}
 along a given trajectory after determining the opacity $\kappa_{\lambda}$ in each grid cell as a local function of temperature and pressure (see Section~\ref{sec:opacity sources} for a discussion of our opacity sources), where $ds$ is the line-of-sight path length through an individual grid cell. The wavelength at which $\kappa_{\lambda}$ is evaluated is adjusted depending on the line-of-sight velocity of the grid cell $v_{\rm LOS}$, as per
\begin{equation}
\begin{split}
    \lambda = \lambda_0\bigg{(}1 - \frac{v_{\rm LOS}}{c}\bigg{)} \\
     v_{\rm LOS} = -u\sin(\phi) - v\cos(\phi) 
    \sin(\theta) \\- \Omega(R_{\rm P} + z)\sin(\phi)\cos(\theta)
\end{split}
\end{equation}
for rest wavelength $\lambda_0$, east-west velocity $u$, north-south velocity $v$, latitude $\theta$, longitude $\phi$, altitude $z$, speed of light $c$, and planetary angular rotation speed $\Omega$. These Doppler shifts assume that the orbital velocity of the planet has already been entirely accounted for; we address potential consequences of this assumption with respect to WASP-76b in Section~\ref{sec:eccentricity}. As described in \cite{kempton2012constraining}, higher-order line-shifting effects such as gravitational redshifting and microturbulent broadening are expected to be negligible.

\begin{figure*}
\centering 
\includegraphics[scale=0.5, trim=20 0 60 0, clip]{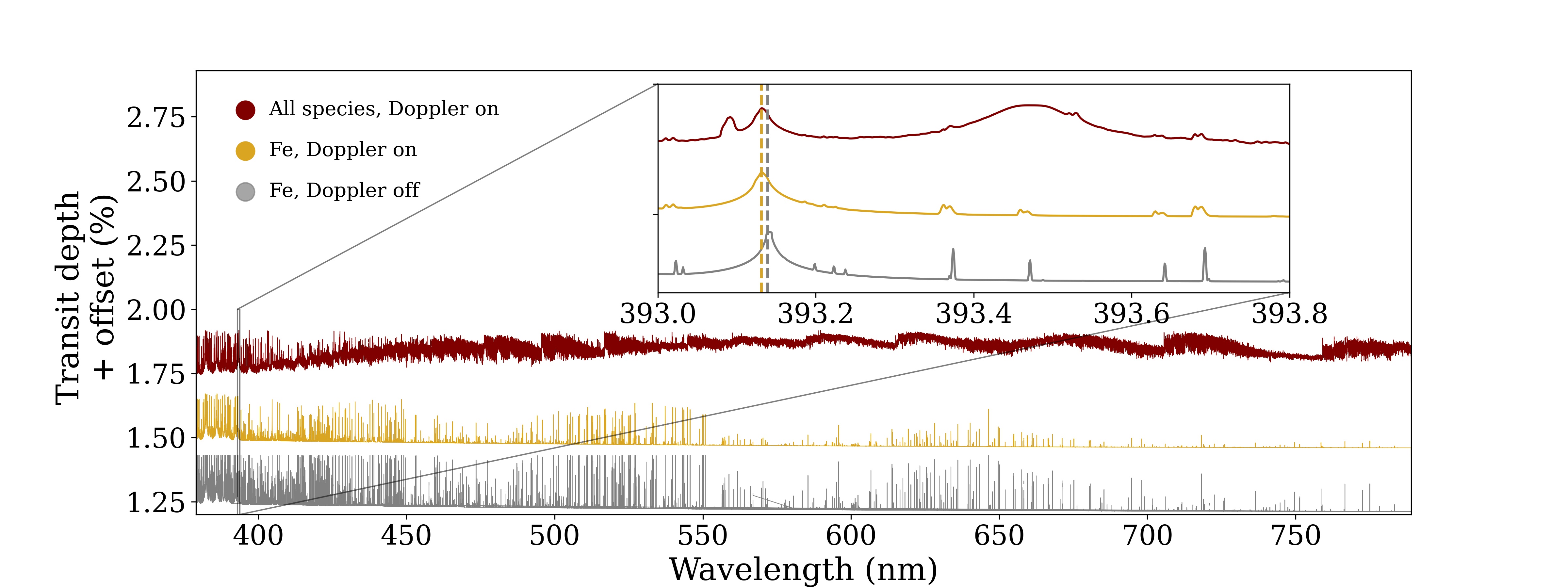}
\caption{Representative model spectra computed in this study. The bottommost, gray spectrum is computed without including Doppler shifts from planetary winds and rotation and includes Fe I and Fe II as the sole opacity sources. The middle, gold spectrum is the same as the bottom, but now including Doppler effects. The topmost, maroon spectrum includes Doppler effects and all opacity sources discussed in Section~\ref{sec:opacity sources}. The inset axis is centered near the Ca II H line (393.4 nm); the vertical dashed lines indicate the peak of a representative Fe feature for the gray and gold spectra, respectively. In the optical range of WASP-76b, the inclusion of Doppler effects broadens and blueshifts strong lines, such as those depicted.}
\label{fig: spectrum}
\end{figure*}

After calculating the above $\tau_{\lambda}$, we can calculate the intensity $I_{\lambda}$ along each line of sight, assuming absorption-only radiative transfer:
\begin{equation}\label{eq:intensity}
    I_{\lambda} = I_{\lambda, 0}e^{-\tau_{\lambda}} 
\end{equation}
for stellar intensity incident upon a grid cell, $I_{\lambda, 0}$. We assume that the stellar spectrum follows a blackbody distribution for $T = 6329$ K, the effective temperature of WASP-76 \citep{ehrenreich2020nightside}. The total flux transmitted through the atmosphere is arrived at by integrating $I_{\lambda}$ over the sky-projected solid angle of each respective grid cell through which a beam emerges. The wavelength-dependent transit depth $D_{\lambda}$ is then given by 
\begin{equation}\label{eq:transmission}
    D_{\lambda} =  \frac{F_{out}-F_{in}}{F_{out}},
\end{equation}
where $F_{out}$ is the out-of transit flux (i.e.\ the stellar flux) and $F_{in}$ is the in-transit flux, accounting for both the blockage of stellar light by the optically thick core of the planet and the transmitted flux through the atmosphere.

Our spectral simulations are computed over the \mbox{379 -- 789~nm} range to coincide with the ESPRESSO data from \cite{ehrenreich2020nightside}, with a total of 298,766 individual wavelength points corresponding to a resolution of $R \approx 4 \times 10^5$. After computation, our spectra are convolved to the native ESPRESSO resolution in the singleHR mode \citep[$R \approx 138,000$;][]{pepe2013espresso}. Because of very large file sizes and RAM requirements associated with the substantial number of wavelength points, we generate the spectra in discrete wavelength ``chunks'' on a high-performance computing cluster and then stitch the entire spectrum together once the full calculation is complete.

\subsubsection{Opacity sources}\label{sec:opacity sources}

Because WASP-76b is an ultra-hot Jupiter observed at optical wavelengths, we compile a set of opacities that is appropriate for these conditions. The atomic species that we utilize are Fe~I, Fe~II, Mg~I, Mn~I, Na~I, Ca~II, Ti, Li~I, and K~I; and our molecular species are TiO, VO, OH, and $\rm H_2O$. Rayleigh scattering from H$_2$, H, and He, as well as $\rm H^-$ bound-free and free-free absorption are included as continuum opacity sources. Finally, collision-induced opacity from collisional pairs of $\rm H_2$-$\rm H_2$, $\rm H_2$-H, $\rm H_2$-He, and H-He is considered. This list of opacity sources includes all of the species with reported detections in WASP-76b optical spectra from \citet{ehrenreich2020nightside}, \citet{tabernero2021espresso}, and \citet{fu2020hubble}. We do not include ZrO, which was searched for in WASP-76b by \cite{tabernero2021espresso} but resulted in a non-detection. In modeling WASP-76b, we generate two sets of transmission spectra, one with Fe (I and II) and continuum opacity sources only, and another with the full set of molecular, atomic, and continuum absorbers (see Figure \ref{fig: spectrum} and Section~\ref{sec:iron-only}).

All opacities are generated on a temperature-pressure grid spanning from \mbox{500 -- 5000~K} and \mbox{1$\times 10^{-6}$ -- 1$\times 10^{3}$~bar}, covering the range of the WASP-76b GCM output. We initially calculate opacities on an evenly spaced wavenumber grid, from a wavenumber of 0 to 30,000~$\rm cm^{-1}$ with a spacing of 0.01 $\rm cm^{-1}$. These results are then downsampled with the k-table feature of \texttt{HELIOS} \citep{malik2017helios}.  

We use the newest available open-source line lists in constructing our molecular opacities: VO \citep[VOMYT;][]{mckemmish2016exomol}, TiO \citep[TOTO;][]{mckemmish2019exomol}, OH \citep[MoLLIST;][]{tennyson2016exomol}, $\rm H_2O$ \citep[POKAZATEL;][]{polyansky2018exomol}. Atomic opacities are drawn from the Kurucz database \citep{kurucz1995kurucz}. In selecting our opacities, we make use of the most abundant isotopologue for each species. Our atomic sources do not include pressure broadening --- only Doppler (thermal) and natural broadening.\footnote{The Doppler effects from planetary rotation and winds dominate over the broadening introduced by pressure, thermal, and quantum effects here, because the signal for high-resolution spectra originates from high-altitude, low-pressure regions of planetary atmospheres. Furthermore, our science case is primarily concerned with the location of line centers, not the exact nature of line wings. Therefore, we anticipate that the exclusion of pressure broadening here does not significantly impact our final spectra.} From these line lists, we calculate the relevant opacities with the GPU-enabled \texttt{HELIOS-K} code \citep{helios-k}.

Opacities from individual species are combined by a weighting of their respective mixing ratios, under the assumption that the atmosphere of WASP-76b resides in a state of chemical equilibrium. To determine the chemical equilibrium mixing ratios, we employ the \texttt{FastChem} model \citep{stock2018fastchem}, which accounts for gas-phase chemistry only. In Section~\ref{subsec:condensation}, we describe how we treat condensation for the purposes of our modeling.

\subsubsection{Phase-dependent transmission spectra}\label{sec:phase treatment}

As in \citet{flowers2019high}, to compare our model transmission spectra directly against the \citet{ehrenreich2020nightside} results, we must calculate transmission spectra as a function of orbital phase throughout the duration of transit. To account for orbital phase dependencies, we apply the following procedure:

\begin{enumerate}
    \item \textit{Account for phase-dependent back-lighting of the planet (i.e.\ stellar limb-darkening effects).} At different points of its transit, the planet will occult regions of its host star of varying brightness. Furthermore, at a fixed orbital phase, different regions of the planet's limb will be back-lit by varying intensities of stellar light. Similarly to \citet{flowers2019high}, we calculate the normalized stellar intensity at the center of each cell of the 2-D projected planetary grid produced by our GCM at each modeled orbital phase of the planet. We use the quadratic limb-darkening coefficients reported by \cite{ehrenreich2020nightside} to establish the stellar center-to-limb intensity profile, and we take into account the 89.623$^\circ$ orbital inclination of WASP-76b \citep{ehrenreich2020nightside} to determine where the planet resides on the stellar disk as a function of its orbital phase. We make the assumption of constant impact parameter $b$ over the course of transit.\footnote{In reality, a planet on an inclined orbit will not have a constant $b$ over the entire duration of transit; rather, the planet's distance from the stellar equator will be decreased at ingress and egress, reaching its maximum at center of transit. Our tests reveal that the relative error induced by the constant $b$ assumption is on the order of 4\% in distance, which results in a change on the order of 1 m/s at the blueshift level (see Section~\ref{sec:cross-correlation}). Hence, our $b$ treatment is justified.} This procedure allows us to calculate a back-lighting factor $f$, which ranges from 0~to~1, effectively replacing the constant $I_{\lambda, 0}$ from Equation~\ref{eq:intensity} with a variable $I_{\lambda, 0} \times f(\theta', z, \varphi)$, for a given orbital phase $\varphi$ and 2-D projected polar angle $\theta'$.
    \item \textit{Account for the decreasing of the continuum by interpolating a light curve produced by the \texttt{batman} code \citep{kreidberg2015batman}.} Step 1 ensures that less light is transmitted through the planet's atmosphere than a uniform stellar disk would emit. Step 2 further enforces that the inner, optically thick core of the planet is simulated crossing a limb-darkened star, as opposed to a star of uniform brightness.
    \item \textit{Account for the planet's rotation over the course of transit.} Because the planet is continually rotating as it travels across the face of its host star, we must transform the GCM coordinate system so that the correct observer-facing hemisphere is modeled at each instance during transit. For simplicity, we assume zero obliquity, which allows us to calculate the coordinate transform simply by assigning a linear offset to each planetary longitude; i.e., $\phi_{\rm rotated} = \phi + \varphi$.

\end{enumerate}

In choosing phases at which to evaluate our models, we more densely sample ingress and egress because the transmission signal quickly varies in strength and velocity during these times. The sampling that we employ is: 
\begin{equation}
    d\varphi =
  \begin{cases}
    0.416\degr, & -15.94\degr \leq \varphi \leq -8.25\degr \\
    1.39\degr, & -8.25\degr \leq \varphi \leq 6.0\degr \\
    0.416\degr, & 6.0\degr \leq \varphi \leq 15.94\degr
  \end{cases}
\end{equation}
where $\varphi_{\rm max} = 15.94 \degr$ represents the phase at which the planet's trailing limb no longer occults the stellar disk. This procedure results in 50 total orbital phases being explicitly modeled with our radiative transfer post-processing code.

\subsubsection{Condensation and clouds \label{subsec:condensation}}

The large blueshifts observed by \citet{ehrenreich2020nightside} have been attributed to iron condensation on the nightside and cooler ``morning" limb of WASP-76b. To confront this hypothesis within the framework of our transmission spectrum modeling, we account for clouds and condensation in the following ways.

To incorporate the effects of condensation out of the gas phase, we utilize the iron condensation curve for a solar composition mixture computed in \cite{mbarek2016clouds}. At each cell in our GCM-produced atmosphere, we interpolate the condensation curve at the cell's pressure; if the temperature in the cell is lower than the interpolated condensation temperature, then we remove iron from the total opacity calculation at that location by setting its opacity contribution to zero. This approach maximizes the effect of Fe condensation, as in reality homogeneous nucleation of Fe is inefficient \citep{gao2021universal}. This step effectively treats the physics considered in the \citet{ehrenreich2020nightside} toy model, which only accounted for gas phase removal as the process responsible for the large net blueshifts in WASP-76b's transmission spectrum.

In our full-species modeling of WASP-76b's transmission spectrum, we allow other species to condense out of the atmosphere following a similar procedure. We condense Ca (I and II) following the $\rm Ca_2SiO_4$ condensation curve, Ti and TiO following $\rm CaTiO_3$, K following $\rm KCl$, Na following $\rm Na_2S$, Mn following MnS, and Cr following $\rm Cr_2O_3$. Our prescription of complete removal implies full rainout from the gas phase once the species-limiting condensate forms.

Condensation not only causes gas-phase depletion but can also result in the formation of an optically thick cloud made up of liquid droplets or solid particles. Such a cloud layer would block transmission of stellar light through the atmosphere in the location where the cloud forms and could therefore also be the main driver of the ``missing'' iron absorption in the receding limb of WASP-76b's atmosphere --- a scenario that has not yet been explored in the literature for this planet. To account for such cloud layers, 
our models further contain a post-processing implementation of gray, optically thick clouds. These clouds, which are self-consistent with our radiative transfer (but not with our GCMs), are placed such that they span up to several ($1-10$) scale heights above the condensation curve of iron, forsterite ($\rm Mg_2SiO_4$), or aluminum oxide ($\rm Al_2O_3$). 

Fe is selected because it is the species that is putatively condensing in WASP-76b's atmosphere; $\rm Mg_2SiO_4$ is modeled because per the microphysics model of \cite{gao2020aerosol}, silicate clouds are chemically favored over iron ones and appear to be a defining feature in the transmission spectra of giant planets with comparable irradiation to WASP-76b; and $\rm Al_2O_3$ is considered because it is the highest-temperature condensate predicted in hot Jupiter atmospheres \citep{mbarek2016clouds}.

\begin{figure*}
\centering
\includegraphics[scale=0.525, trim=120 0 0 0, clip]{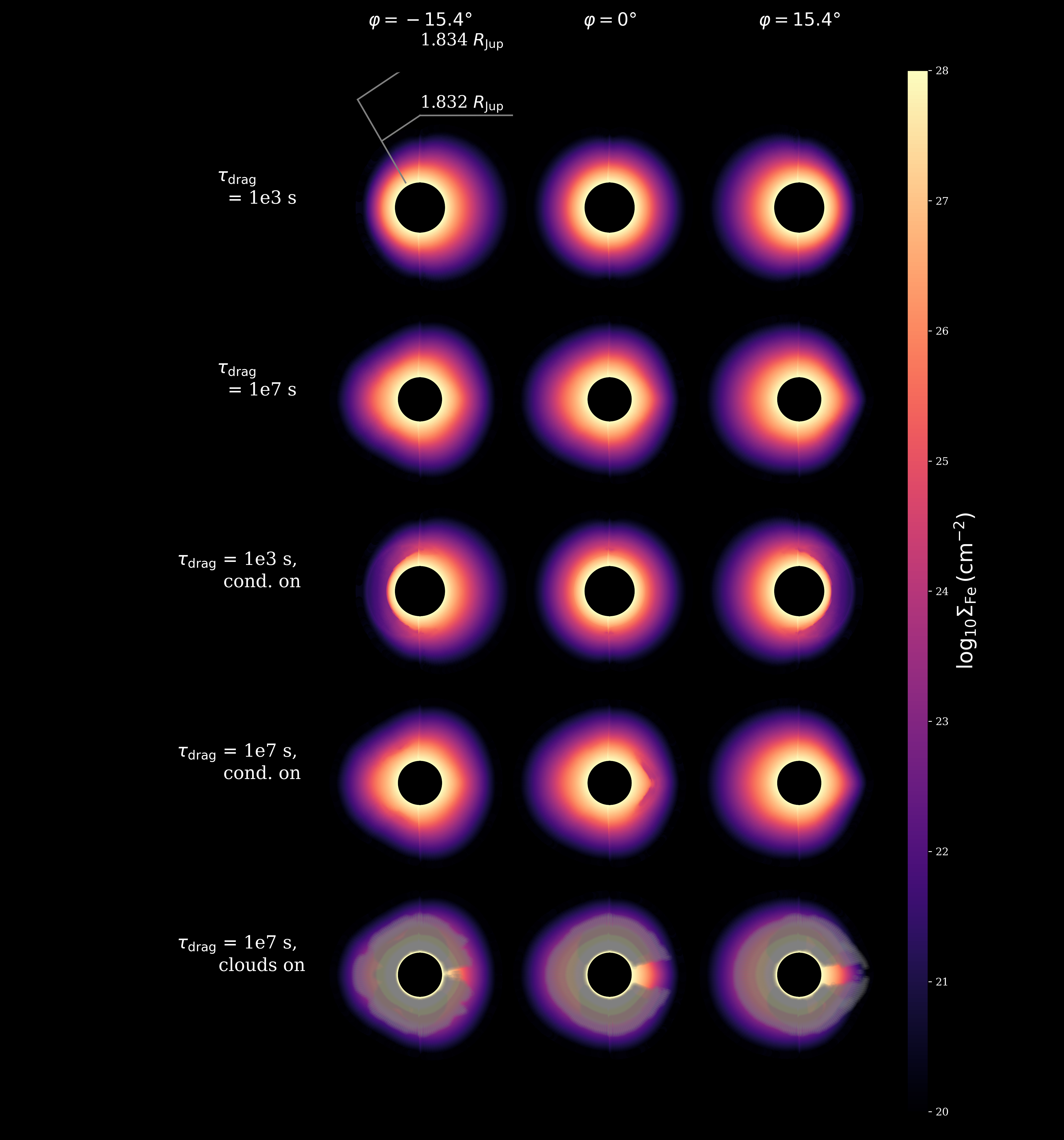}
\caption{Column-integrated surface density of Fe abundance ($\Sigma_{\rm Fe}$) of WASP-76b. Each figure column corresponds to an orbital phase (left to right: ingress, center of transit, and egress), and each row corresponds to a model run (top to bottom: strong drag, weak drag, strong drag with condensation, weak drag with condensation, weak drag with Fe clouds). All models shown have a deep RCB. One scale height clouds are drawn with a darker gray, and 10 scale height clouds are drawn with a lighter gray.  (In practice, the latter do not achieve a full thickness of 10 scale heights because the upper atmosphere of WASP-76b presents conditions incompatible with iron condensation --- see text for more information.) The strong drag case is more symmetric, especially at center of transit, than the weak drag case.} 
\label{fig:abund_maps}
\end{figure*}

Notably, our implemented clouds never fully reach the limiting-case thickness of 10 scale heights. Because our temperature-pressure profiles tend to cross condensation curves twice (Figure~\ref{fig: t-p profs}), clouds cannot exist in the uppermost regions of the atmosphere --- the upper boundary of the cloud is (indirectly) set by the pressure at which the atmospheric inversion or isothermal region occurs. This effect can be seen in the bottom row of Figure~\ref{fig:abund_maps}, in which the ``10 scale height'' clouds do not extend much past the ``1 scale height'' clouds. Hence, while our clouds can exist up to 10 scale heights in thickness (provided they fulfill their criterion with respect to a given condensation curve), in practice they are limited by the temperature/pressure structure of the atmosphere.

Similarly to our condensation treatment, our cloud effect is modeled \textit{post-hoc}, in that the radiative effects of cloud formation and the resultant changes to opacities are not accounted for natively in the GCM (see Section~\ref{sec:model_limitations} for a discussion of this limitation). Furthermore, our simple cloud model does not account for the wavelength-dependent scattering and/or absorption from cloud particles --- a simplification that is justified because cloud opacities are typically smoothly varying with wavelength, in contrast with the sharply wavelength-dependent atomic and molecular features that are primarily responsible for the high-resolution transmission spectrum signal in cross-correlation. Additionally, cloud particle sizes are represented by broad distributions, and no clouds in a planetary atmosphere are characterized by a single particle size \citep[e.g.,][]{pont2013prevalence}.

\subsection{Suite of models \label{sec:model description}}
The suite of post-processed spectral models explored in this work is laid out in Table~\ref{table:parameters}.  We use the following terminology to refer to our various model implementations.
By ``condensation on,'' we mean that we are applying the rainout prescription described in Section~\ref{subsec:condensation}. ``Fe-only'' models refer to those in which the only non-continuum opacity source is gas-phase iron. Our ``full-species'' models include the full set of opacities listed in Section~\ref{sec:opacity sources}.

Furthermore, when referring to cloud inclusion, ``Fe clouds on'' implies the placement of an optically thick cloud above the Fe condensation curve with a thickness equivalent to a specified number of scale heights. Similarly, ``$\rm Mg_2SiO_4$'' or ``$\rm Al_2O_3$ clouds on'' refer to the same, albeit above the $\rm Mg_2SiO_4$ or $\rm Al_2O_3$ condensation curves, respectively. The inclusion of clouds in our models is decoupled from our condensation procedure, allowing us to model these effects together or separately.

The majority of our work holds WASP-76b on a circular orbit (i.e., $e$ = 0). We investigate the effect of eccentric orbits in Section~\ref{sec:eccentricity}.

\section{Analysis}\label{analysis}
\subsection{Cross-correlation}\label{sec:cross-correlation}
With our model spectra in hand, our next step is to determine the information that can be recovered from these spectra. For high spectral resolution observations of exoplanet atmospheres, signal recovery has typically been accomplished via a cross-correlation analysis \citep[e.g.,][]{snellen2010orbital,birkby2013detection,brogi2016rotation}. This procedure has an advantage over the fitting of individual spectral lines (Section~\ref{ind-lines}) in that it leverages a key aspect of high-resolution data --- a forest of weak lines that together can produce a high-SNR detection of a chemical species. To mimic the data analysis process undertaken by \citet{ehrenreich2020nightside}, thereby allowing us to compare our models directly to their results, we implement a cross-correlation procedure into our modeling routine.

\begin{figure}
\centering
\includegraphics[scale=0.35, trim=20 0 20 0, clip]{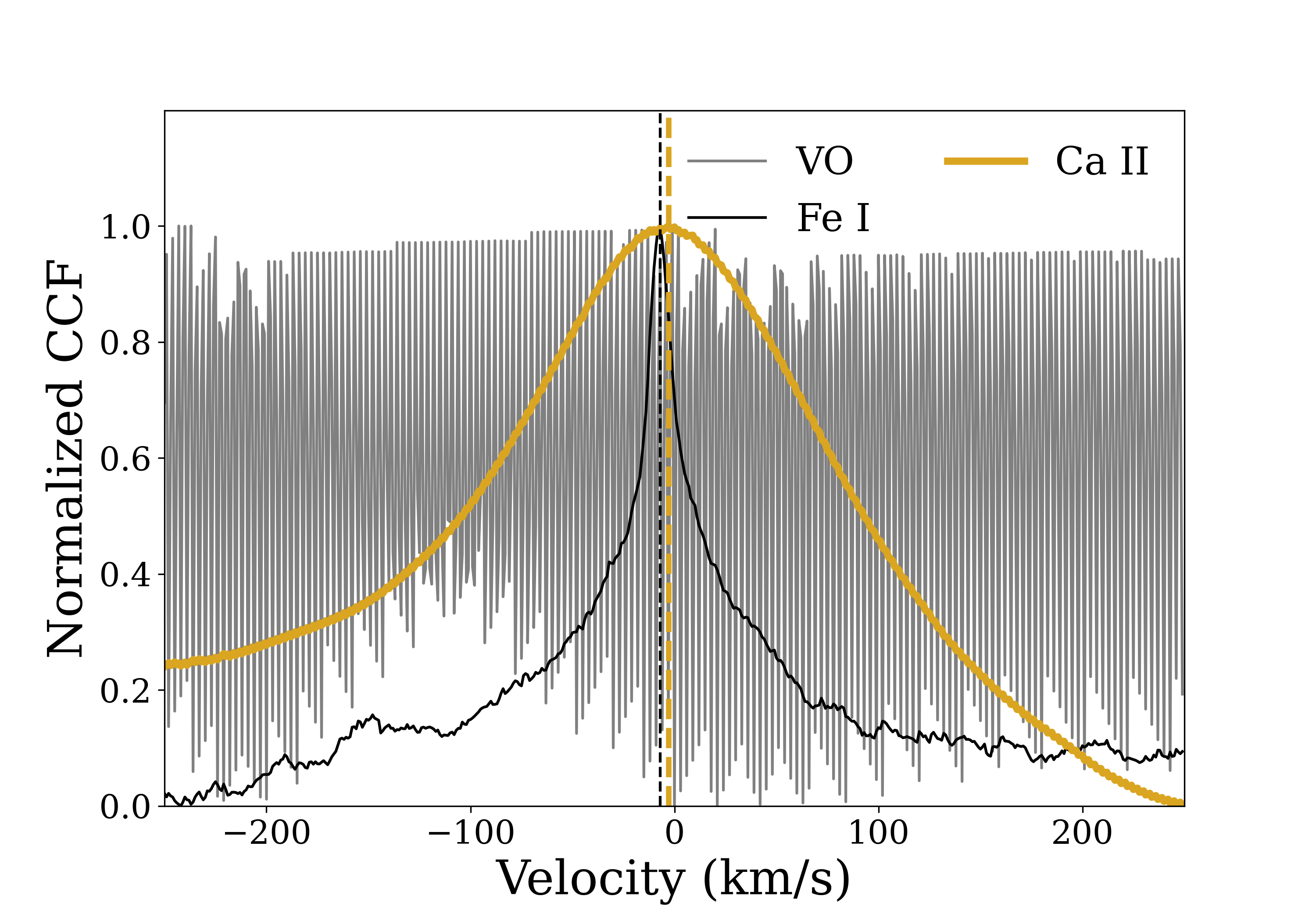}
\caption{Normalized cross-correlation functions of VO (gray), Fe I (black), and Ca II (gold). Note that while the CCFs of Fe I and Ca II have clearly defined peaks (indicated by vertical dashed lines), the VO CCF is very noisy, with no discernible peak value. Most of our computed CCFs are similarly bimodal in quality.}
\label{fig:ccf}
\end{figure}

Our cross-correlation function (CCF) $c$ is computed by combining our data, $x$, with a mask or template, $T$, at a given velocity, $v$, as follows \citep[e.g.,][]{baranne1996elodie, pepe2002coralie,allart2017search,hoeijmakers2019spectral}:
\begin{equation}
    c(v) = \sum_{i=0}^{N} x_iT_i(v)
\end{equation}
where the mask or template is Doppler-shifted by velocity $v$ and interpolated onto the wavelength grid of $x$ for summing. This calculation produces a map of cross-correlation against velocity (Figure~\ref{fig:ccf}); the location of the peak of this distribution is reported as the net Doppler shift of $x$. We Doppler-shift $T$ between $-250$~km\,s$^{-1}$ and 250~km\,s$^{-1}$ in steps of 1~km\,s$^{-1}$.

We first employ a model-on-model cross-correlation. To do so, we construct as our template a simplified, ``Doppler-off'' spectrum, which is a model otherwise equivalent to the simulated data but with no Doppler shifts applied in the radiative transfer calculation. This template is then convolved down to the resolution of ESPRESSO using a Gaussian kernel of appropriate width. We then perform the previously described cross-correlation procedure.

Once we perform our cross-correlation, we fit a Gaussian to our CCF with the SciPy \texttt{curve\_fit} routine. The peak value of the fitted Gaussian is then reported as the velocity shift of a given spectrum.

For CCFs with well-characterized peaks, we perform a second, narrower Gaussian fit in a window 6~km\,s$^{-1}$ wide centered on the peak identified by the previous step. Doing so ensures that asymmetries in the CCF far from the peak do not influence the final determination of the peak value. Our fitting process allows us to report CCF peak values to precision higher than the 1 km\,s$^{-1}$ sampling of our CCFs; limited testing of sampling CCFs at higher velocity resolution yielded results consistent with our chosen sampling. CCFs that are not well-characterized (e.g., VO; Figure~\ref{fig:ccf}) are not considered in further analysis.

\subsection{Individual line-fitting} \label{ind-lines}
For certain strong lines, we perform individual fitting to determine Doppler shifts of those lines. This process is an important complement to the cross-correlation procedure described above, and it allows us to compare our forward models directly to the observational results of \citet{tabernero2021espresso}, who focused on specific strong lines of a range of chemical species. 
Individual-line Doppler shifts are also scientifically interesting because lines of different strengths probe different altitudes of the planet's atmosphere, and hence different portions of the planet's wind field and temperature structure \citep[e.g.,][]{kempton2012constraining}. 

To fit individual lines, we fit a Gaussian profile to a narrow spectral window centered on the line in our center-of-transit spectrum. The corresponding Doppler shift ($v_{\rm Doppler}$) is calculated by comparing the wavelength at the maximum of the fit Gaussian ($\lambda_{\rm max, fit}$) to the rest wavelength ($\lambda_0$):

\begin{equation}
    v_{\rm Doppler} = c\bigg{(}1 - \frac{\lambda_{\rm max, fit}}{\lambda_0}\bigg{)}
\end{equation}

For ease of comparison to the \cite{tabernero2021espresso} data, we fit the same lines examined by that study. Furthermore, we adopt their method for fitting weak lines: As an intermediate approach between cross-correlation of many lines and fitting a single line, \cite{tabernero2021espresso} combine multiple (maximum 5) weak lines in velocity space and fit them jointly to produce a stronger signal. As with that work, we apply this approach for Fe, Mn, and a subset of Mg lines.

This line-fitting method allows us to probe both the Doppler shifts of individual lines and their depths --- which in turn establish the altitude at which these lines become optically thick.

\section{Results}\label{results}
\subsection{Cross-correlation}
\subsubsection{Fe-only}\label{sec:iron-only}
For our first set of results, we perform our analysis on our iron-only transmission spectra.
\begin{figure*}
    \centering
    \includegraphics[scale=0.4]{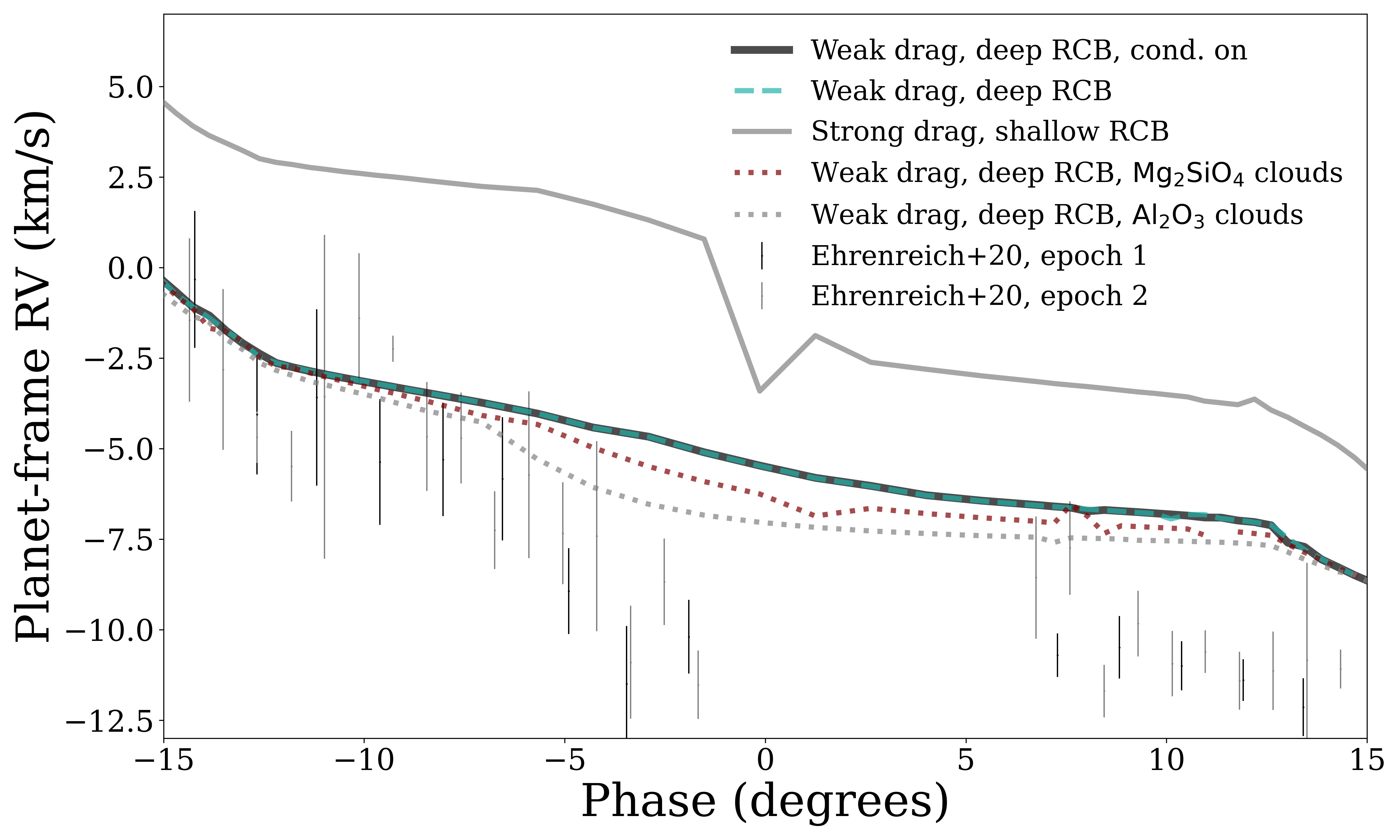}
    \caption{Representative planet-frame Doppler shifts as a function of orbital phase for our Fe-only models, as computed by our cross-correlations of forward-modeled spectra. Overplotted are data from the \cite{ehrenreich2020nightside} WASP-76b observations and their corresponding 1$\sigma$ error bars (corresponding to the top panel of that work's Extended Data Figure 7). Our worst-fitting model consistent with our GCM has strong drag ($\tau_{\rm drag} = 1e3$ s) and a shallow RCB (solid gray line). Conversely, our best-fitting model consistent with our GCM has weak drag ($\tau_{\rm drag} = 1e7$ s) and a deep RCB (dashed teal line). Finally, our best-fitting model with 1 scale height clouds inclusive of condensation effects has weak drag, a deep RCB, and $\rm Al_2O_3$ composition (dotted gray line).}
    \label{fig:phase_results_iron_only}
\end{figure*}

\begin{figure*}
\centering
\includegraphics[scale=0.52, trim=40 0 60 0, clip]{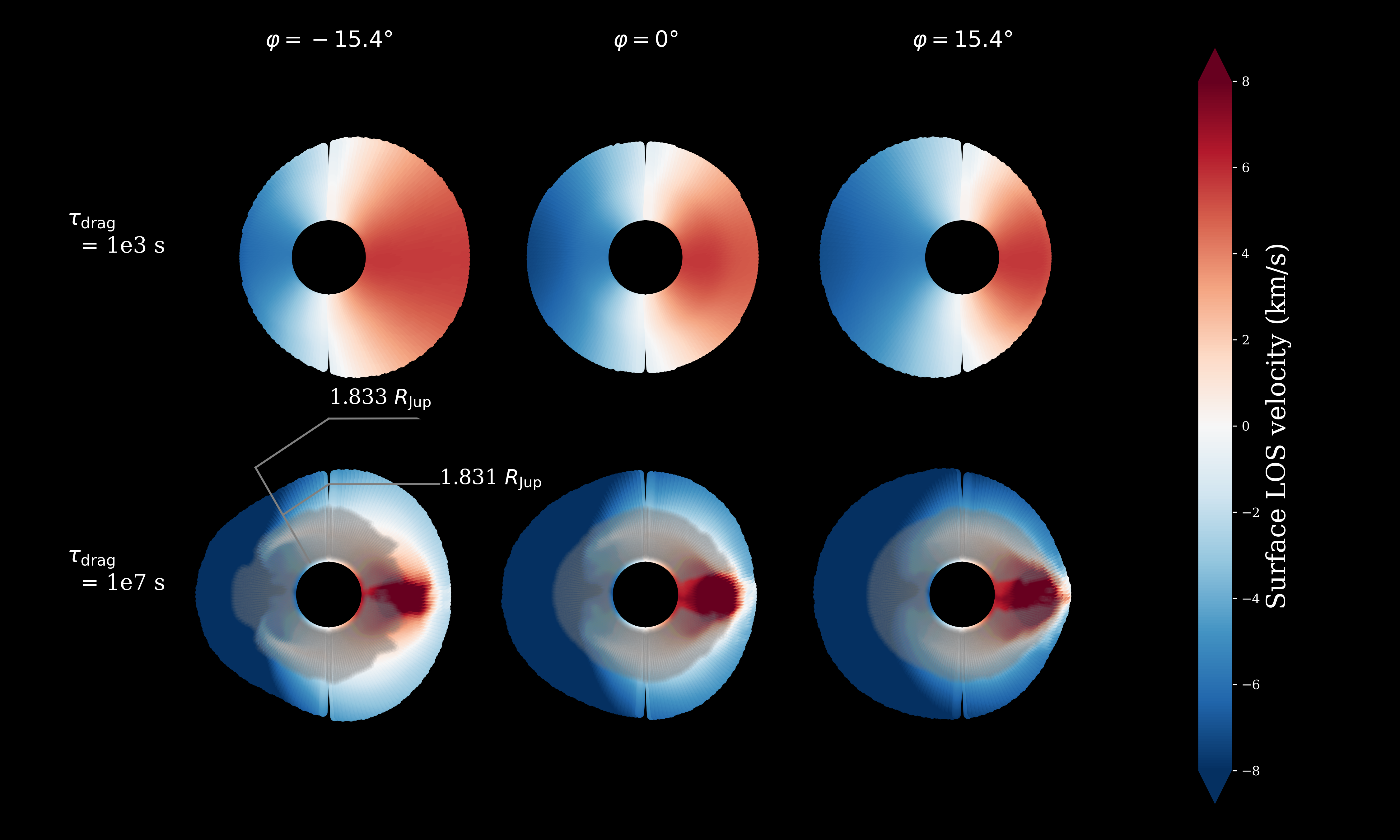}
\caption{Terminator line-of-sight velocity maps of our deep RCB WASP-76b GCMs, including contributions from both winds and rotation, plotted in the style of Figure~\ref{fig:temp_map} (two drag timescales and three phases). Our (Fe) modeled clouds are overplotted as a gray band in the bottom row, for the 10 scale height implementation. Our clouds, as opposed to our wind field, is not a terminator slice, but rather a depiction of every (projected) sightline that intersects a cloud in the full 3D geometry. As in maps depicting other quantities, the line-of-sight velocities are more symmetric for the high-drag case, while they are more asymmetric for the weak-drag case. Stronger blueshifting is further present at egress than ingress for the weak-drag case.}
\label{fig:wind_map}
\end{figure*}
We find that controlling the drag timescale over our explored parameter range has the strongest first-order effect on Doppler shift over all phases (Figure \ref{fig:phase_results_iron_only}), with the weak drag models providing a much better match to the \citet{ehrenreich2020nightside} data than the strong drag models. Decreasing the drag timescale in an atmosphere is equivalent to making drag processes more efficient. In the Helmholtz decomposition framework of \cite{hammond2021rotational}, about half the day-night heat redistribution in a fiducial hot Jupiter atmosphere is driven by substellar-antistellar flow; shorter drag timescales would serve to slow the speed of this flow, reducing heat transport. The remainder of the day-night heat redistribution, per \cite{hammond2021rotational}, is due to rotational flow --- e.g., the superrotating equatorial jet. The speed of this jet would similarly suffer directly from a shorter drag timescale. Additionally, per \cite{showman2012doppler}, stronger drag also reduces the pumping of eddy angular momentum onto the dayside equator because drag damps the propagation of planet-scale wave patterns, further decreasing the jet strength. In sum, increased drag reduces longitudinal asymmetries and slows wind speeds, both of which reduce net Doppler shifts in transmission spectra.

As a result of the aforementioned effects, our high-drag models have more homogenized limbs, whereas low-drag models have limbs that are more heterogeneous, with a visible equatorial jet structure (Figures~\ref{fig:temp_map}, \ref{fig:abund_maps}, \& \ref{fig:wind_map}). These features can be seen in our simulated Doppler shifts (Figure~\ref{fig:phase_results_iron_only}). The strong drag case has Doppler shifts that are roughly symmetric about the center of transit, with only a slight ($<2$~km\,s$^{-1}$) blueshift at mid-transit. Moreover, the maximum and minimum of the Doppler shift are both near $\pm$ 4--5~km\,s$^{-1}$, which is close to the equatorial rotational velocity of WASP-76b ($\pm$~5.3~km\,s$^{-1}$). These features are expected for a solely rotationally broadened profile --- which would be predicted for the high-drag models, which have weaker winds. Additionally, a clear prediction from Figure~\ref{fig:wind_map} is that the strong drag case is redshifted at ingress and blueshifted at egress --- as seen in Figure~\ref{fig:phase_results_iron_only}.

Similarly, the Fe abundance map (Figure~\ref{fig:abund_maps}) combined with the wind map (Figure~\ref{fig:wind_map}) explains the behavior of the weak drag case in Figure~\ref{fig:phase_results_iron_only}. Namely, the Fe abundance is asymmetrically distributed, preferentially on the eastern limb, where the strongest blueshifts for this model are. Therefore, the model produces Doppler shifts that are not evenly distributed around 0~km\,s$^{-1}$ across phase and are more strongly blueshifted, which is more consistent with the \citet{ehrenreich2020nightside} observations.

Including iron condensation removes the gas-phase iron in the cooler regions of the atmosphere (Figure~\ref{fig:abund_maps}). As can be inferred from our temperature-pressure structure (Figure~\ref{fig: t-p profs}), condensation is primarily confined to the nightside in all our GCMs. Especially in our weak drag models, condensation is less prominent at the equator, whereas it is more prominent at the mid-latitudes and poles. The effect of condensation tends to be marginal, dependent on cloud placement, and not necessarily linear, reflective of the complexities in the 3-D thermal structure of the atmosphere.

\begin{figure*}
    \centering
    \includegraphics[scale=0.7]{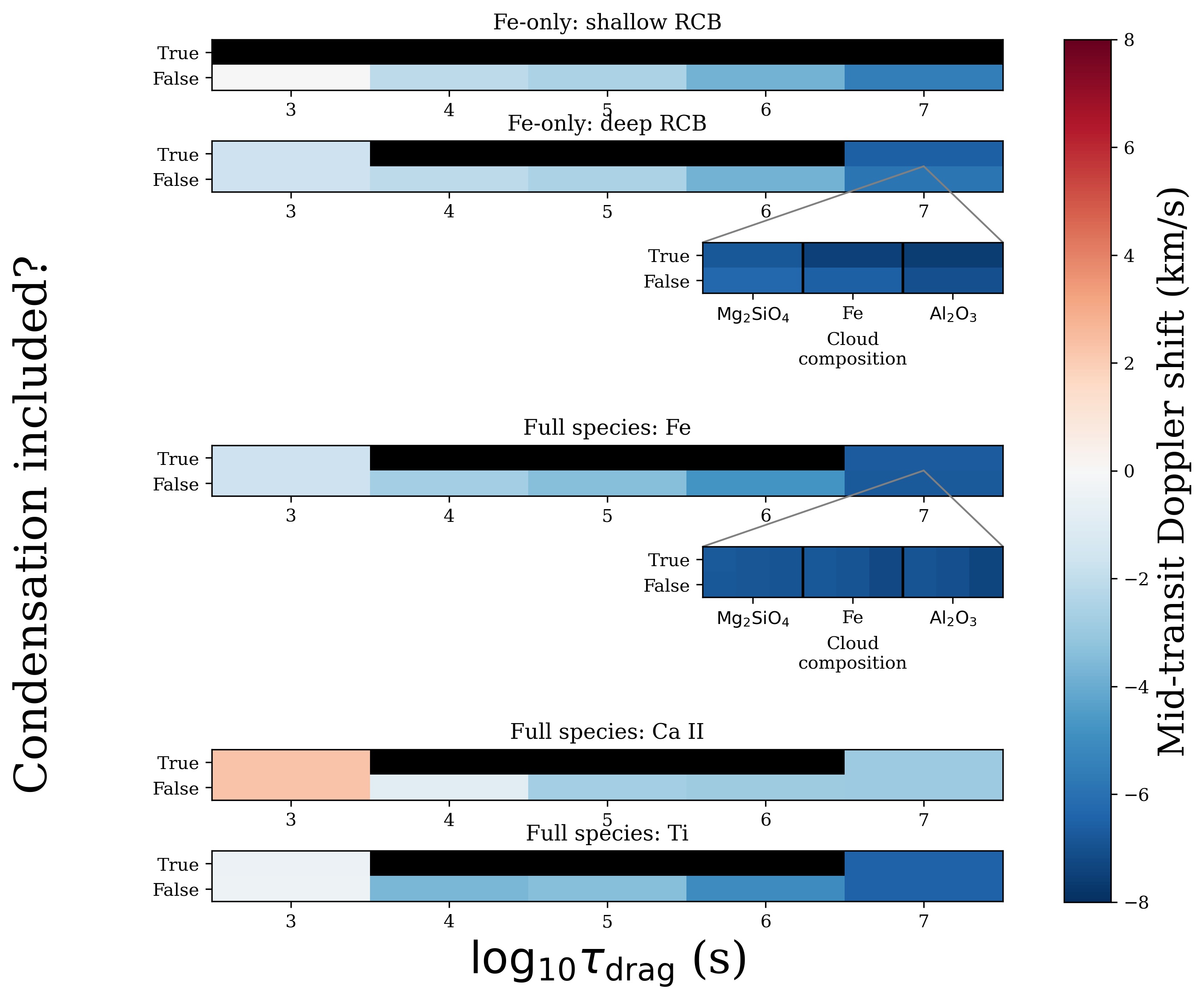}
    \caption{Center-of-transit Doppler shifts for all models explored in this work. From left to right, we show the effect of increasing the drag timescale of our GCM (i.e., decreasing the effect of drag within the atmosphere). Within a given row, we include gas-phase condensation for some models (``True" vs.\ ``False"), as described in Section~\ref{subsec:condensation}. For the weakest drag timescales, we also explore the effect of placing optically thick clouds above different species' condensation curves ($\rm Mg_2SiO_4$, Fe, and $\rm Al_2O_3$), as described in Section~\ref{subsec:condensation}. \addressresponse{Cells representing models that are not computed are filled in black.} The upper sets of models presented here only includes iron as a spectral opacity source (``Fe-only''), whereas the lower sets of models include all gas opacity sources discussed in Section~\ref{sec:opacity sources} (``Full Species''). For our full species models, we explore the effect of cross-correlating against different single-opacity template spectra (e.g., ``Full species: Fe'' convolves against a full species spectrum against an Fe template) and increasing our clouds' scale heights (left to right: 1, 5, 10 scale heights). The strongest blueshift is detected for the all-species model / Fe cross-correlation mask with condensation on, and 10 scale height thick $\rm Al_2O_3$ clouds. Aside from the top row, all models are computed with a deep RCB. For reference, the observed transit-averaged Doppler shift identified by \cite{tabernero2021espresso} via Fe cross-correlation is $-8.25$ $\pm$ 0.25~km\,s$^{-1}$ for the first analyzed transit (T1) and $-8.75$ $\pm$ 0.56~km\,s$^{-1}$ for the second analyzed transit (T2). Line analysis by \cite{tabernero2021espresso} in T1 yielded Doppler shifts of 4.1 $\pm$ 5.1~km\,s$^{-1}$ and $-4.4$ $\pm$ 2.5~km\,s$^{-1}$ for the Ca II H\&K lines, respectively. T2 analysis produced Doppler shifts of 1.0 $\pm$ 3.0~km\,s$^{-1}$ and $-2.1$ $\pm$ 1.9~km\,s$^{-1}$ for these lines.}
    \label{fig:all_ccfs}
\end{figure*}

We further find that our deep RCB models are a better fit to the data, generating stronger net blueshifts. A shallow RCB is expected to create a more homogeneous atmosphere, because convective motions quickly mix entropy across large spatial scales, reducing day-night temperature contrasts on deep isobars relative to the deep RCB case. Atmospheric circulation in (ultra-)hot Jupiters is driven and supported by day-night pressure gradients \citep[e.g.,][]{showman2002atmospheric}, so the smaller day-night contrasts in the shallow RCB case cause winds to be slower. These slower winds in turn decrease east-west limb asymmetries as well as the strength of the blueshift signal (as shown in Figures ~\ref{fig:phase_results_iron_only} \& \ref{fig:all_ccfs}).

We find that the presence of optically thick clouds, as opposed to gas-phase iron condensation, also strongly and consistently increases the measured blueshift (Figure~\ref{fig:phase_results_iron_only}). This is because clouds will preferentially form on the cooler, receding (eastern) limb, hence blocking its contribution to the transmission spectrum; these clouds would lessen the redshift contribution of the receding limb, thereby boosting the measured blueshift.

As seen in Figure \ref{fig: t-p profs}, $\rm Al_2O_3$ condenses out at hotter temperatures than Fe. Hence, placement of clouds above the $\rm Al_2O_3$ condensation curve as opposed to the Fe condensation curve results in a cloud deck situated in warmer regions of the atmosphere. The effect of this change at the blueshift level is highly nonlinear, as it depends somewhat strongly on the exact placement (and thickness) of the cloud. Notably, though, the optically thick clouds produce a ``jump'' in blueshift near a phase of $-7$\degr\ that the other models cannot (Figure \ref{fig:phase_results_iron_only}) --- a distinct feature of the \cite{ehrenreich2020nightside} data. While this jump does not extend to the exact observed magnitude, it appears that an $\rm Al_2O_3$ cloud deck (or a similarly positioned cloud of arbitrary composition) provides the type of spatial inhomogeneity required to reproduce the \cite{ehrenreich2020nightside} data in broad strokes.

Interestingly, although the deep RCB / strong drag GCM would have much more condensation in its atmosphere (Figure \ref{fig: t-p profs}), the deep RCB / weak drag GCM has much stronger blueshifts (Figure \ref{fig:all_ccfs}). It thus appears that the importance of the drag timescale and faster winds leading to more east-west asymmetry ``wins out'' over the existence of condensation alone.

Our net center-of-transit Doppler shifts are summarized in Figure~\ref{fig:all_ccfs} for all of the model permutations explored. In total, we find that the following effects produce increased blueshifts in WASP-76b's atmosphere when iron is considered alone: weak drag, deep RCBs, and optically thick clouds composed of Al$_2$O$_3$, Mg$_2$SiO$_4$, or Fe. Gas-phase iron condensation increases net blueshifts in some cases, but (contrary to the \citealt{ehrenreich2020nightside} interpretation) actually decreases net blueshifts in others.

By toggling our back-lighting effect, we determine that accurate treatment of limb-darkening (see Section~\ref{sec:phase treatment}) has a median effect over the course of transit on the order of 0.1~km\,s$^{-1}$. Considering the mean errorbar size on the \cite{ehrenreich2020nightside} blueshift data (Figure~\ref{fig:phase_results_iron_only}), it appears that although including this effect improves physical consistency, it does not have strong (or even currently detectable) effects at the Doppler shift level.

\subsubsection{Full species}
For our next set of model runs, we include all of the opacity sources described in Section~\ref{sec:opacity sources} and cross-correlate these ``full species'' models against Doppler-off templates that include one chemical species at a time. We are able to identify peaks in the CCFs at most in-transit phases for Fe, Ti, Mn`11, Ca II, and Cr, indicating that these species are present at an observable level in our modeled spectra.

\begin{figure*}
\gridline{\fig{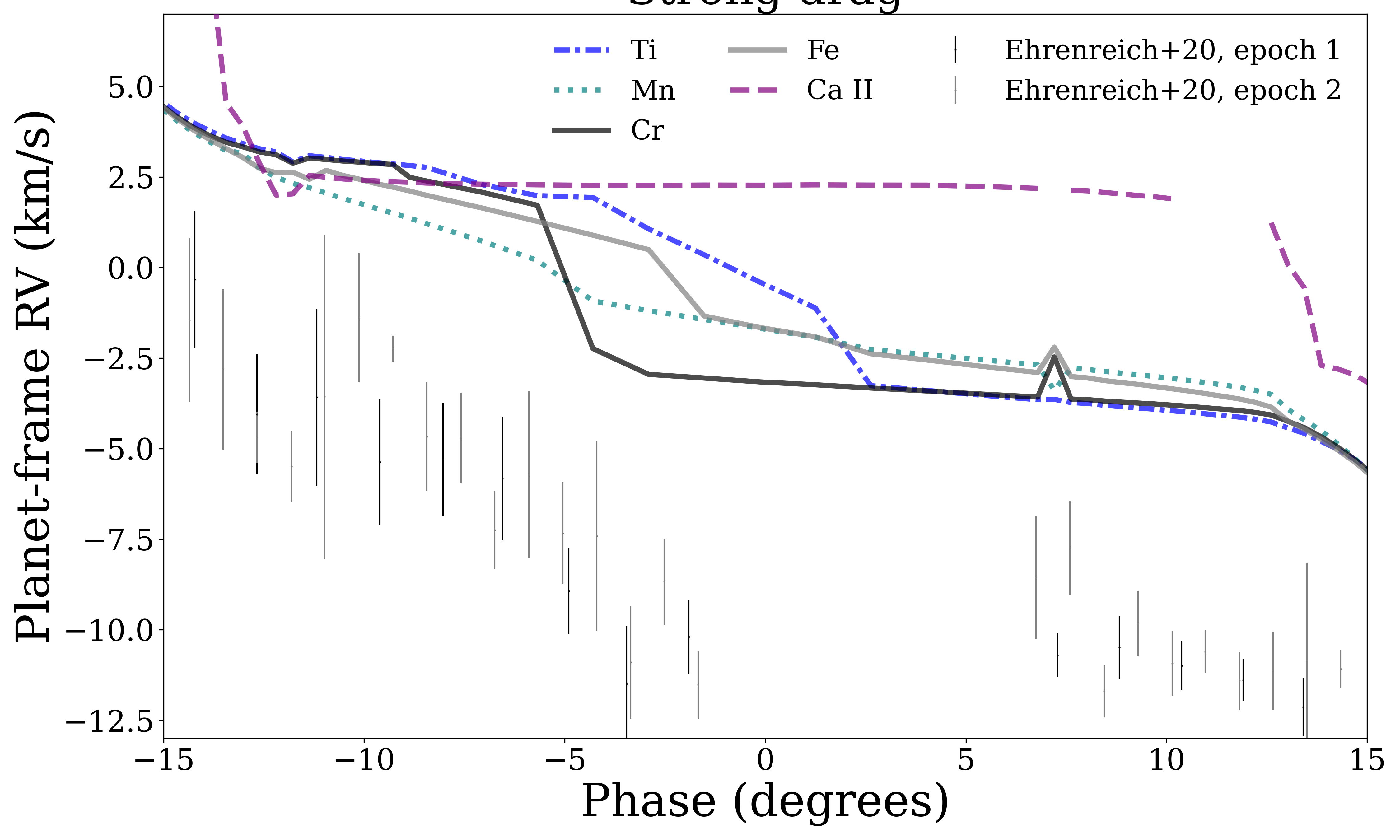}{0.47\textwidth}{(a)}
          \fig{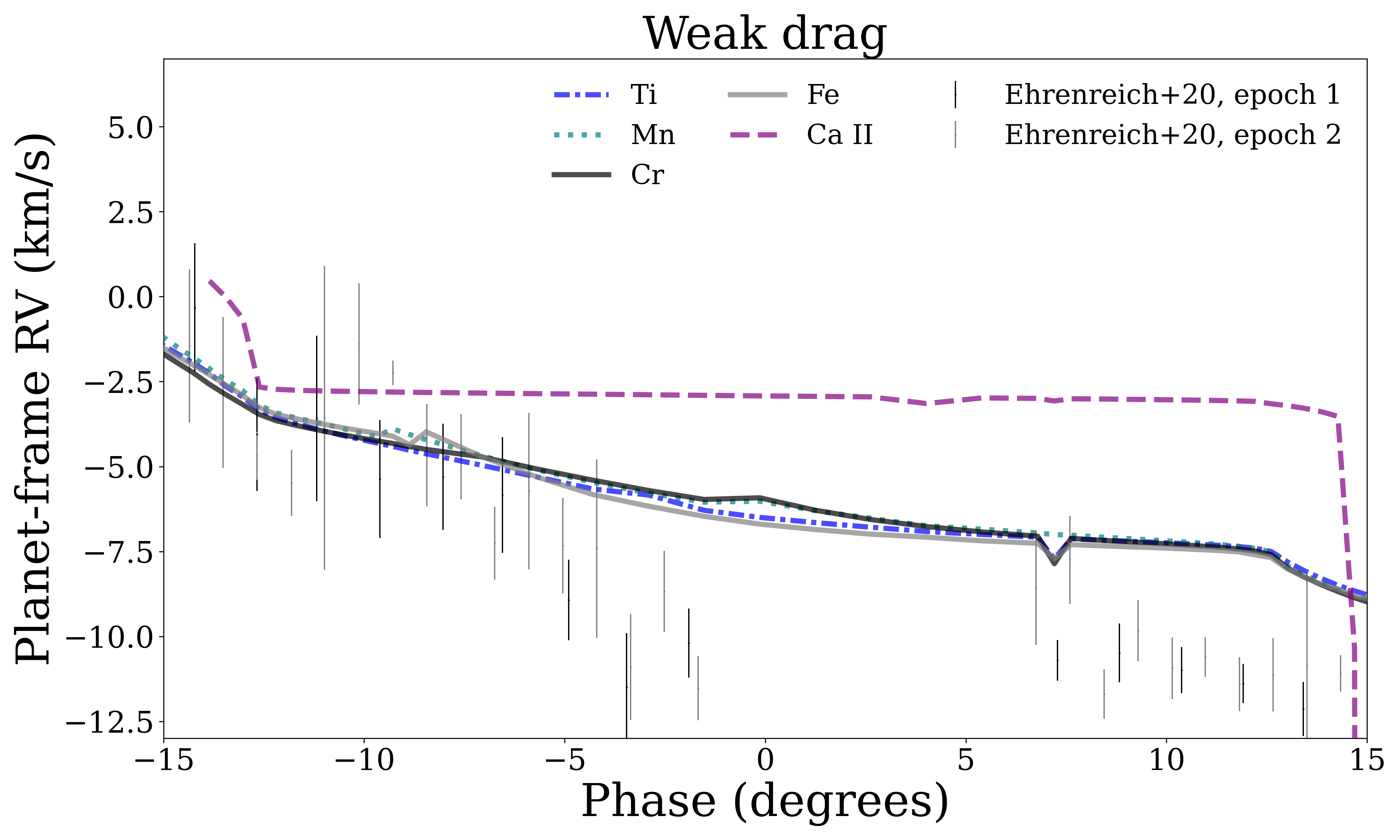}{0.47\textwidth}{(b)}
          }
\gridline{\fig{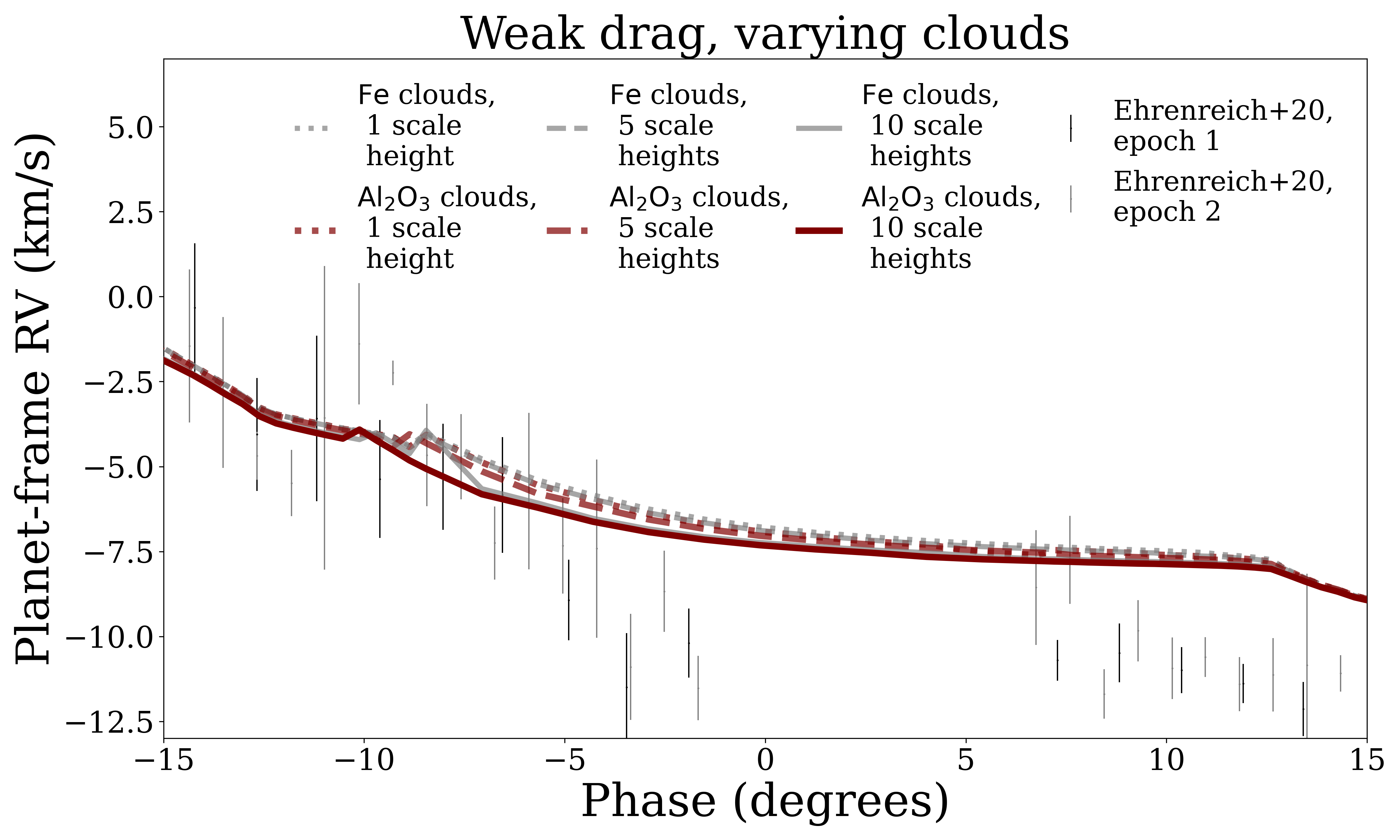}{0.47\textwidth}{(c)}
            \fig{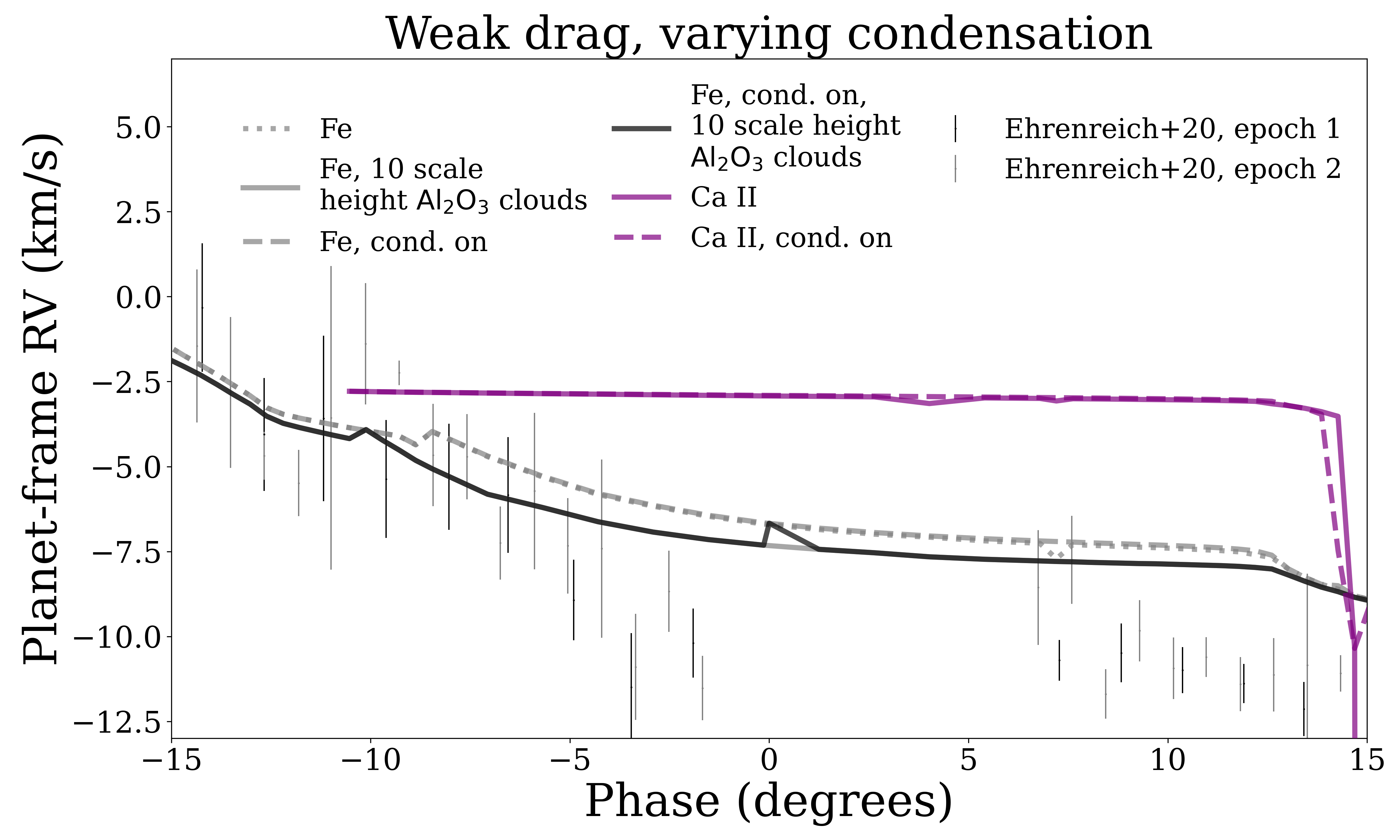}{0.47\textwidth}{(d)}
          }
\caption{Phase-resolved CCFs of our spectra including all opacity sources. Subplot (a) is the strong drag case, subplot (b) is the weak drag case, subplot (c) showcases the weak drag case for varying cloud treatments (only convolving with an Fe template), and subplot (d) compares the condensation on/off cases. Most species with detectable CCFs follow the Doppler trend of Fe I, with the notable exception of Ca II. Note that the CCF of Ca II is too noisy to fit during ingress in subplot (c). As in the Fe-only models, introducing thick clouds to our full-species spectrum reproduces the sharp jump in Doppler shift noted by \cite{ehrenreich2020nightside} and \cite{kesseli2021confirmation}, and the condensation \addressresponse{effect} alters the fit for condensed species depending on whether clouds are included; non-condensed species are unaffected by condensation treatments. Our best-fitting model (up to 10 scale height $\rm Al_2O_3$ clouds, weak drag, condensation on, deep RCB) is shown in black in panel (d).}
\label{fig:all_species_ccf}
\end{figure*}

Similarly to the Fe-only models, we find that spectrally post-processed, unaltered GCMs cannot reproduce the \cite{ehrenreich2020nightside} piecewise blueshift trend without added modifications such as condensation or cloud formation (Figure \ref{fig:all_species_ccf}, panels (a) and (b)). Most of the detected species tend to follow the trend of Fe, especially in the weak drag case. The clear exception is Ca II. It exhibits much weaker blueshifts than the other species in the weak drag case, and is mostly redshifted over the course of transit in the strong drag case. Interestingly, its Doppler shift remains mostly constant out to egress in the weak drag case, whereas it experiences a sharper blueshift in egress in the strong drag case. This trend implies that in the weak drag case, the Ca II signature is proportionally stronger in the approaching limb, so as the receding limb exits the stellar disk in egress, the Doppler signal remains largely unchanged until the approaching limb exits as well. In contrast, in the strong drag case, it appears that the Ca II signal is more evenly distributed on both limbs, so when the receding limb begins to exit the stellar disk, the total contribution of Ca II signal shifts to the approaching limb, rapidly increasing the measured blueshift over egress due to the rotational component.

These characteristics can be physically motivated by considering the temperature structure of the atmosphere. The abundance of Ca II in equilibrium chemistry is strongly dependent on thermal ionization, and hence temperature. Any temperature asymmetries, then, will be reflected in asymmetries in the abundance profile of Ca II. Per Figure \ref{fig:temp_map}, our weak drag case is much hotter on the approaching limb, because the lack of strong drag is conducive to the formation of a superrotating equatorial jet that advects heat from the  planet's hotspot, causing a thermal offset in the direction of the approaching limb. Therefore, in the weak drag case, the approaching limb will be more abundant in Ca II than the receding limb. In the strong drag case, however, the lack of a jet and hotspot offset imposes limb homogeneity, giving each limb relatively equal weight in the Ca II signal. See Section \ref{sec:hydrostatic} for a further discussion of Ca II and its potential non-hydrostatic behavior in observations.

Adding clouds to our models (Figure \ref{fig:all_species_ccf}, panel (c)) has the same effect identified in the Fe-only models: stronger blueshifts and a trend that better matches the \cite{ehrenreich2020nightside} data for clouds with higher condensation temperatures.

Notably, the cloud thickness seems to strongly control the jump in blueshift prior to center of transit --- the greater the maximum thickness of the cloud, the earlier in phase the jump occurs. The best-fitting model in this grid includes clouds up to 10 scale heights above the $\rm Al_2O_3$ condensation curve (corresponding to a 0.7~mbar cloud top on the western limb and a 1.1~mbar cloud top on the eastern limb). A cloud top up to 10 scale heights above the $\rm Fe$ condensation curve performs nearly, but not quite as well. Even in the $\rm Al_2O_3$ cloud case, though, the fit to the data is not perfect; namely, the magnitude of our Doppler shift is a few km\,s$^{-1}$ short of the observed data at egress.  (Figure~\ref{fig:all_species_ccf}, panel (c)).

Contrary to the cloudless case, including gas-phase condensation when clouds are also included increases the measured blueshift (Figures~\ref{fig:all_ccfs} \& \ref{fig:all_species_ccf}). This effect is often very slight; in the 10 scale height, $\rm Al_2O_3$ cloud case, including condensation increases blueshift by 6~m\,s$^{-1}$ at center of transit.

 The overall best fit to the \citet{ehrenreich2020nightside} data from our suite of models is obtained for the case of low drag ($\tau_{\mathrm{drag}} = 10^7$), a deep RCB, a 10 scale height, optically thick $\rm Al_2O_3$ cloud, and gas-phase iron condensation. Even this model, however, is not perfect. For instance, this model fails to match the egress blueshift magnitude of the \citet{ehrenreich2020nightside} data (Figure~\ref{fig:phase_results_iron_only}). We discuss these points in further detail in Section~\ref{discussion}.

\begin{figure}
\centering
\includegraphics[scale=0.26, trim=10 0 10 0, clip]{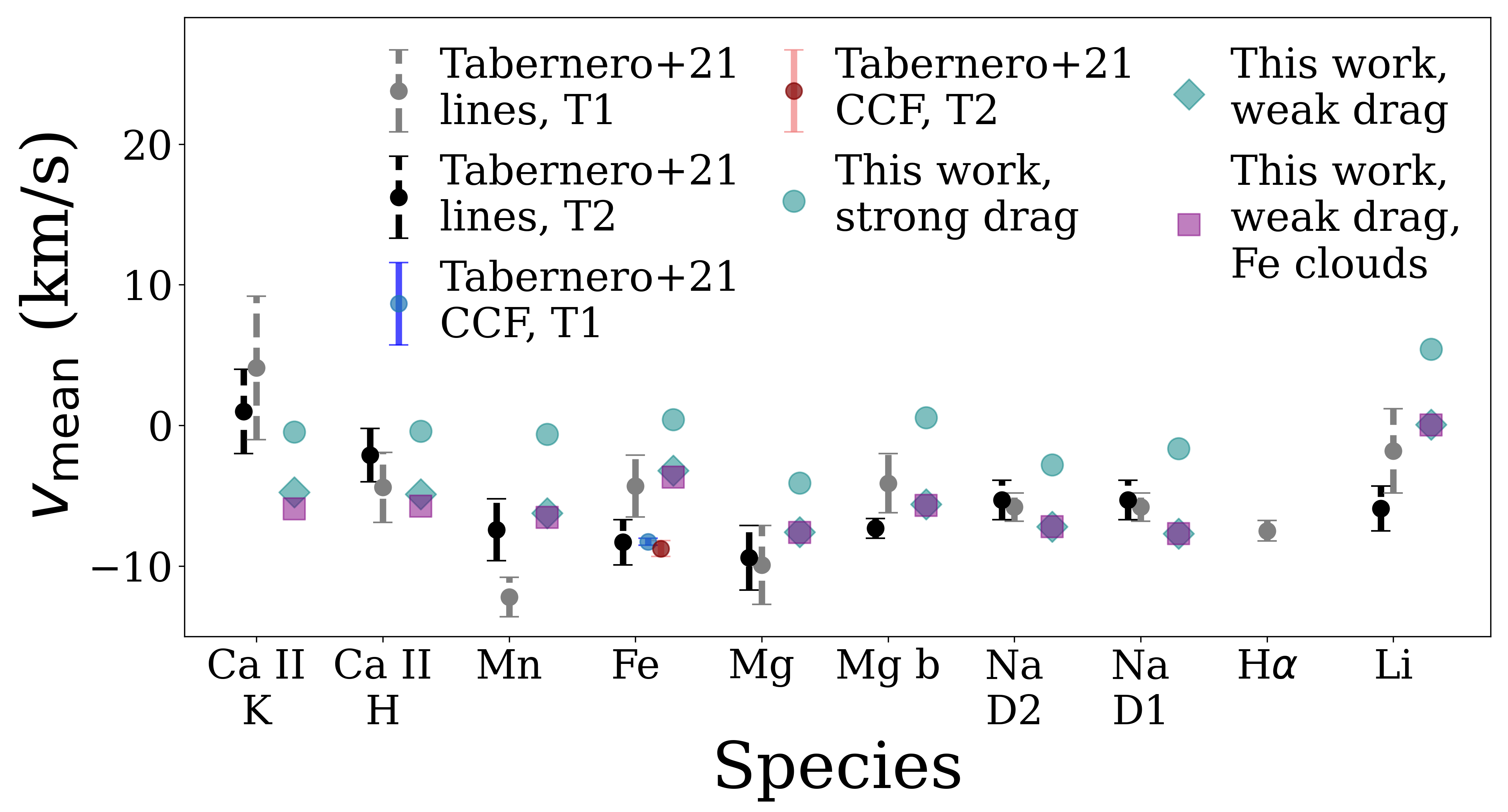}
\caption{Doppler shifts investigated by \cite{tabernero2021espresso} in comparison with the lines from the forward models in our study. Including 10 scale height, gray Fe clouds in our model (purple squares) blueshifts our lines to varying degrees across species. The weak drag models are slightly preferred over the strong drag case. Although endmembers of drag (teal circle for strong drag, teal diamond for weak drag) are plotted for clarity, intermediate drag seems to be slightly more consistent with the observed Doppler shift data overall.}
\label{fig:line_speeds}
\end{figure}

\begin{figure}
\centering
\includegraphics[scale=0.26, trim=10 0 10 0, clip]{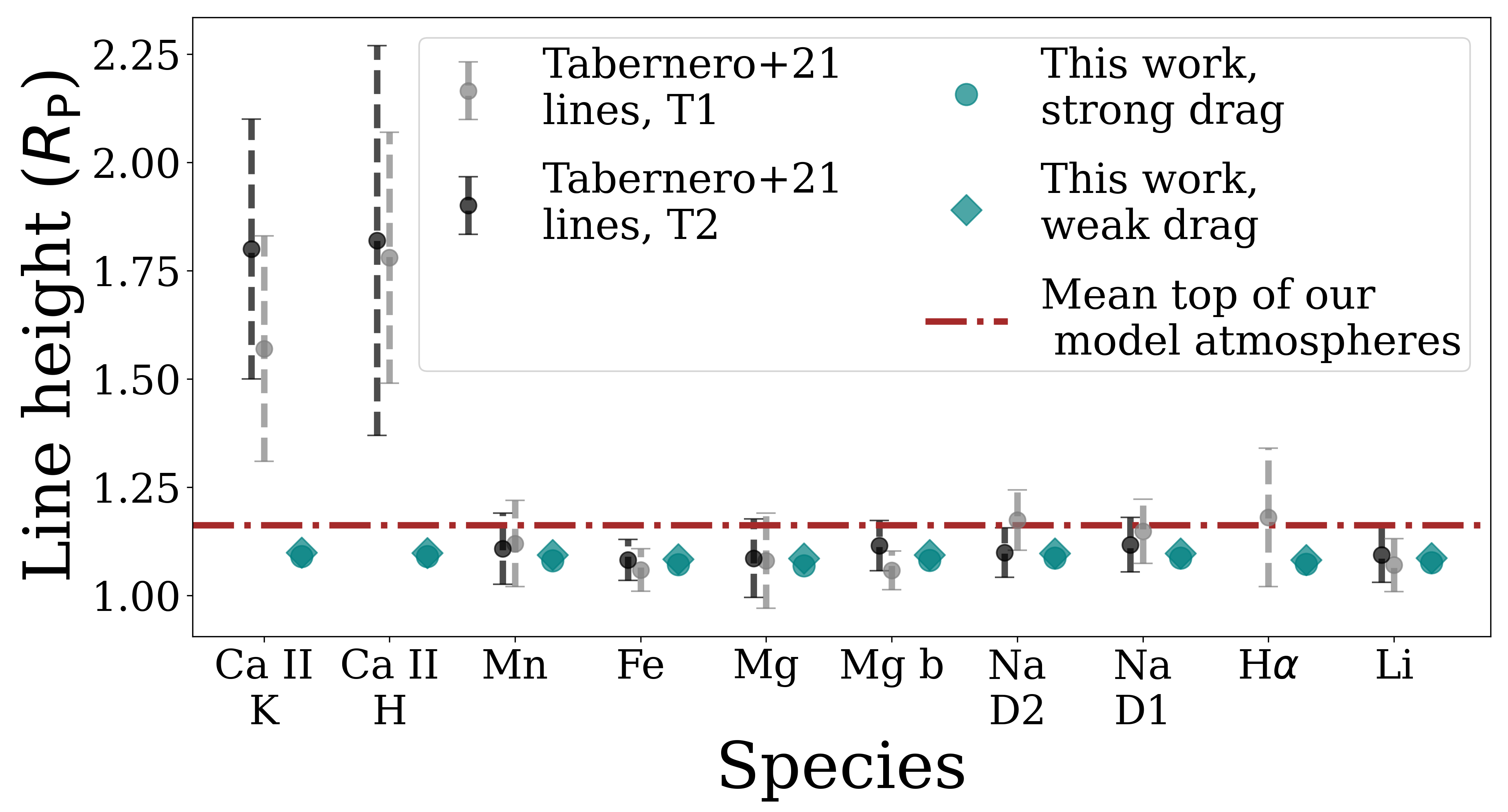}
\caption{Height of lines investigated by \cite{tabernero2021espresso} in comparison with the lines from representative forward models in our study. Our strong drag case is plotted with teal circles, whereas our weak drag case is plotted with teal diamonds. Most of our lines are consistent with the \cite{tabernero2021espresso} line heights --- with the exception of the Ca II lines, which have observed heights consistent with an escaping atmosphere that extends beyond our model domain.}
\label{fig:line_heights}
\end{figure}

\subsection{Line-fitting}

Overall, we are able to detect and fit nearly all of the individual absorption lines explored by \cite{tabernero2021espresso}. We additionally find that our line Doppler shifts are all consistent within 2$\sigma$ with the \cite{tabernero2021espresso} data (Figure~\ref{fig:line_speeds}), and nearly all our line heights are consistent within 1$\sigma$ for at least one of the \cite{tabernero2021espresso} transits (Figure~\ref{fig:line_heights}). 

As also seen in our cross-correlation analysis, the lines from our weak-drag models have stronger blueshifts than our strong drag models. Condensation has a negligible effect on the Doppler shifts of the non-condensed species and a small effect on the Doppler shifts of the condensed species. Introducing clouds has the expected impact of increasing blueshift across species (Figure \ref{fig:line_speeds}). Yet some species (Ca II, Mn I) are more affected by the inclusion of clouds than others (Na I, Li I). This is not surprising, as different species are expected to be abundant in different temperature-pressure regimes, and they would thus be blocked by our pressure-dependent cloud treatment to differing extents. 

Broadly, as with the \citet{ehrenreich2020nightside} data, the \cite{tabernero2021espresso} Doppler shift data favor the weakest drag case over the strongest, though they cannot discriminate between different condensation cases. The line height data cannot favor one model over another, given their error bars. Even so, both data sets can be used to demonstrate an overall degree of consistency of our models with the available data.

All of the strong lines in our forward models fall within the altitude range of 1.078 $R_{\rm P}$ to 1.095 $R_{\rm P}$ (Figure~\ref{fig:line_heights}). For most species, this is commensurate with the \cite{tabernero2021espresso} line height data. Notably, however, our models significantly fail to reproduce the observed Ca II line strengths (see Section~\ref{sec:compare_tabernero}).

\section{Discussion}\label{discussion}
\subsection{Comparison to other work}
\subsubsection{Comparison to the \cite{ehrenreich2020nightside} ``Toy Model''}\label{toy model}
As shown in Figure \ref{fig:wind_map}, our 3-D, physically self-consistent GCM output wind field is more complex than the substellar-antistellar and rotational wind field described by the \cite{ehrenreich2020nightside} toy model. For instance, blueshifts from winds are not uniform across the entire limb, and there even exists a redshifted annulus on the terminator (bottom row of Figure \ref{fig:wind_map}), which represents return flow from the nightside to the dayside.\footnote{Specifically, we can confirm that this redshifted annulus represents the western flank of the Rossby gyres. \cite{tsai2014three} demonstrate that the shift of a planet's hotspot east of its substellar point is not only a product of heat advection, but is also indicative of a phase shift of planet-scale standing wave (Matsuno-Gill) patterns. Our deep RCB models have stronger day-night contrasts than our shallow RCB models --- hence, the former have stronger equatorial jets and larger phase shifts in their Matsuno-Gill patterns than the latter. The redshifted annulus present in the shallow RCB model (not plotted) can be reproduced in the deep RCB model by rotating the latter into negative phase, which effectively cancels out the additional phase shift.} Furthermore, our GCMs are unable to reproduce the wind speeds required by the \cite{ehrenreich2020nightside} toy model. Our weak drag GCM has an eastern limb average line-of-sight velocity of $-6.5$~km\,s$^{-1}$; this is in excess of the $-5.3$~km\,s$^{-1}$ day-night flow in the toy model. When averaged with the western limb, though, the average line-of-sight velocity reduces to $-2.3$~km\,s$^{-1}$ --- significantly less than the toy model wind field.

A crucial component of the \cite{ehrenreich2020nightside} toy model is an asymmetry in the Fe abundance distribution in the atmosphere of WASP-76b. Broadly, our iron abundance maps (Figure~\ref{fig:abund_maps}) indicate that, regardless of drag timescale, there does exist a projected asymmetry of iron on the eastern limb by the end of transit. The eastern limb is preferentially blueshifted (Figure~\ref{fig:wind_map}), so more gas-phase iron on the eastern limb produces a larger measured blueshift. While this asymmetry is exacerbated by gas-phase condensation of Fe (top two rows vs.\ next two rows of Figure~\ref{fig:abund_maps}), our self-consistent, GCM-based forward models with Fe condensation alone do not readily match either the magnitude or the shape of the \citet{ehrenreich2020nightside} observed blueshifts, and we find that we must invoke additional physics (e.g., clouds) in order to achieve a better fit.

\subsubsection{Comparison to \cite{tabernero2021espresso}}\label{sec:compare_tabernero}

As shown in Figures~\ref{fig:line_speeds} and \ref{fig:line_heights}, our forward models generally provide good agreement with the multiple species detected across the entire ESPRESSO wavelength range by \citet{tabernero2021espresso}. However, we bring up several noteworthy exceptions below.

\cite{tabernero2021espresso} report a detection of $\rm H\alpha$ in one of their two analyzed transits. In our Doppler-off, single-species template models, though, we note that the hydrogen Balmer lines do not protrude above the continuum (comprised of collision-induced absorption, $\rm H^-$ opacity, and Rayleigh scattering). However, when performing a blind search in our full-species spectrum, we find that there does exist an absorption line within the H$\alpha$ spectral region. When compared to the H$\alpha$ rest wavelength, this line (which we identify as a blend of TiO and VO) displays a blueshift relative to rest H$\alpha$ that is 2$\sigma$ consistent with the reported H$\alpha$ data. We propose that either H$\alpha$ was mistakenly identified by \citet{tabernero2021espresso} or that NLTE, non-hydrostatic, and/or non-equilibrium chemistry effects may be boosting the strength of the H$\alpha$ line compared to what is predicted by our GCM post-processing scheme.

Conversely, \cite{tabernero2021espresso} were not able to detect Cr or Ti in the WASP-76b high-resolution transmission spectrum, whereas our chemical equilibrium forward models are readily able to identify these species in cross-correlation. This mismatch may point to unaccounted-for opacity sources that wash out the Cr and Ti features, increased continuum absorption that masks the lines of these species (perhaps excess $\rm H^-$ brought about by photoionization, for example), or non-solar abundances of Cr and Ti.

A notable result of the \cite{tabernero2021espresso} work is the height of their reported Ca II lines. The Ca II H line was detected at a height of 1.57 $R_{\rm P}$, and the Ca II K line was detected at a height of 1.78 $R_{\rm P}$ --- both at high significance and at much higher altitudes than in our models. The error bars on these detections are large enough such that they are still consistent with our model's line heights at 2$\sigma$. However, the consistency of the observed Ca II H\&K line heights with one another supports their fidelity, as their similar line strength implies that they should become optically thick at roughly the same altitude. \addressresponse{Furthermore, the deep absorption of the Ca II triplet near 850~nm detected by two other instruments is supporting evidence for abundant high-altitude Ca II in WASP-76b \citep{casasayas2021carmenes,deibert2021detection}. Finally}, high-altitude Ca II H\&K lines have been detected in other ultra-hot Jupiter atmospheres and have been attributed to hydrodynamic outflows and high-altitude photoionization \citep{yan2019ionized,borsa2021atmospheric}. Therefore, the discrepancy with respect to our models may be linked to the hydrostatic assumption of our GCM (see Section~\ref{sec:hydrostatic}), which only extends up to 1.19 $R_{\rm P}$. Exploring the outflow hypothesis is beyond the scope of this paper, and we encourage further non-hydrostatic investigations of this and similar planets.

\subsubsection{Comparison to \cite{wardenier2021wasp76}}
Similarly to this work, \cite{wardenier2021wasp76} aimed to post-process a 3-D GCM with a radiative transfer code to explore the role of Fe condensation in WASP-76b. Given that they use a different GCM \citep[the nongray SPARC/MITgcm;][]{showman2009atmospheric} and different radiative transfer scheme (Monte Carlo photon-tracking) from those used in this work, comparing basic results of both works serves to validate both approaches.

Our results are generally consistent with the 3-D post-processing results of \cite{wardenier2021wasp76}. Both studies favor weak drag over strong drag models and resort to GCM modifications to reproduce the Doppler shift behavior reported by \cite{ehrenreich2020nightside} and \cite{kesseli2021confirmation}. 

Other details of the modeling approach differ between our two works. For example, \cite{wardenier2021wasp76} are able to use their non-gray GCM to explore the effect of optical opacities on the predicted thermal structure. Our work explores a much larger spectral range than \cite{wardenier2021wasp76}, enabled by our more computationally efficient ray-striking radiative transfer approach. This method allows us to avoid potential biases incurred by only modeling a small portion of the Fe I band structure. (We find this to be up to a 1.5 km\,s$^{-1}$ effect that can further alter the Doppler shift trend over phase, due to Fe lines over a narrow wavelength range only probing a commensurately narrow range of altitudes.) We further model additional opacity sources beyond those considered by \cite{wardenier2021wasp76}.

Notably, \cite{wardenier2021wasp76} were unable to reproduce the post-ingress jump in Doppler shift exhibited by the \cite{ehrenreich2020nightside} and \cite{kesseli2021confirmation} data without artificially restricting the longitude range of iron condensation within their model domain \addressresponse{or removing TiO and VO opacity from their GCM opacities (thus altering the thermal structure of the planet's atmosphere)}. We are able to produce the same jump in our simulations --- effectively muting portions of the gas-phase Fe Doppler shift signal --- but by using a more physically motivated model involving optically thick clouds (Figure~\ref{fig:all_species_ccf}). At no point do we actually change the GCM output (and hence the underlying physics); rather, we use \textit{post-hoc}, painted-on effects to reproduce the Doppler shift observations.

\subsubsection{Consistency with May \& Komacek et~al. 2021}

By comparing the results presented in this paper with those from our companion paper on the Spitzer phase curve of WASP-76b (\linktocite{maykomacek2021spitzer}{May \& Komacek et~al.} \citeyear{maykomacek2021spitzer}), we can attempt to seek out a set of globally consistent models that describe the ensemble of high-resolution and broadband photometric data. We find that models with cold interiors (i.e., a deep RCB) are favored by both sets of observations, as such a model allows for the cold nightside and large phase curve amplitude observed at 4.5 $\mu$m, constrained by the Spitzer study. However, the phase curve data prefer the strong drag ($\tau_{\rm drag} \le 10^4$~s) GCMs, driven largely by the near-zero phase curve offset observed in both Spitzer wavelength channels. This result is in contrast with our own conclusions that the high-resolution data can only be explained by weak drag models, which allow for a significant east-west limb asymmetry.

The mixed success of our joint modeling efforts to consistently describe all observables --- hot spot offsets, phase curve amplitudes, phase-resolved Fe blueshifts, and multi-species blueshifts --- hints that our physical understanding of WASP-76b is as of yet incomplete.

\subsection{Alternative explanations for the anomalous blueshift}
\subsubsection{A small eccentricity}\label{sec:eccentricity}

We have shown that our forward models can reproduce much of the observed Doppler shift trend once optically thick clouds are included (Figure~\ref{fig:phase_results_iron_only}, Figure~\ref{fig:all_species_ccf}, panel (c)). The full magnitude of the observed Doppler shift, however, is not fully accounted for by this approach. The orbit of WASP-76b itself may provide the rest of the solution. 

A Kozai-Lidov mechanism is often invoked in the literature to explain the migration of hot Jupiters to their current-day short-period orbit (for reviews, see \citealt{dawson2018origins} and \addressresponse{\citealt{fortney2021hot}}). Any Kozai-like process, though, would result in vestigial eccentricity, as the mechanism presupposes a high past eccentricity that then enables tidal interactions to reduce the semi-major axis. \addressresponse{Other formation mechanisms also have the potential to drive large eccentricities for Jupiter-mass planets as well --- for instance, migration through the disk into a magnetospheric cavity could do so \citep{debras2021revisiting}. Finally,} hot Jupiters subject to other potential formation channels \citep[even such as in-situ formation; e.g., ][]{batygin2016situ} could have initially low eccentricities excited by secular interactions \addressresponse{\citep[e.g.,][]{wu2011secular}} or an external perturber \citep[e.g.,][]{zakamska2004excitation}.

\begin{figure}
    \centering
    \includegraphics[scale=0.25]{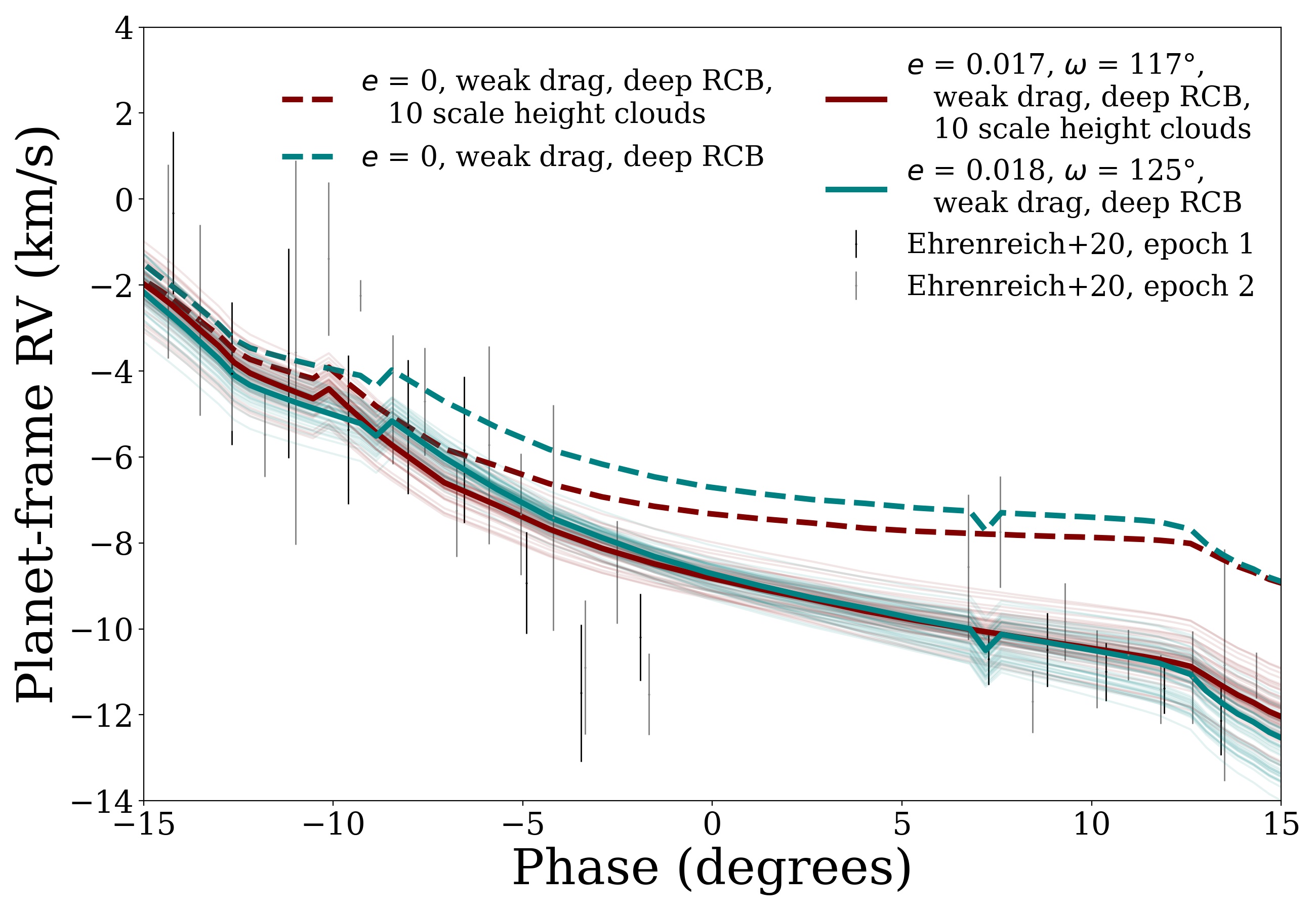}
    \caption{Similar to Figure \ref{fig:phase_results_iron_only}, but now including the effect of a small eccentricity, both with (maroon lines) and without (teal lines) clouds. Circular orbit models are plotted with \addressresponse{dashed} lines; best-fit, eccentricity-inclusive models are plotted with solid lines; and MCMC posterior draws are plotted in faded lines. Our best-fit combination of eccentricity and longitude of periastron (solid teal) is consistent with observations \citep{west2016three, fu2020hubble}. This work's best explanation of the \cite{ehrenreich2020nightside} Doppler shift signature includes 10 scale height $\rm Al_2O_3$ clouds, condensation, and a small eccentricity. The models shown here both include weak drag and a deep RCB.}
    \label{fig:ecc_phase}
\end{figure}
As demonstrated by \cite{montalto2011exoplanet}, the effect of even a small ($e$ = 0.01) unaccounted-for eccentricity can produce velocity shifts for hot Jupiters on the order of km\,s$^{-1}$. With respect to WASP-76b, \cite{west2016three} placed a $3\sigma$ upper bound on eccentricity at 0.05. \cite{fu2020hubble} additionally constrained the planet's eccentricity as $0.016_{-0.011}^{+0.018}$. A nonzero --- and, for our purposes, significant --- eccentricity is therefore certainly within the realm of possibility given the extant data.

In the analyses of \cite{ehrenreich2020nightside} and \cite{tabernero2021espresso}, the eccentricity of the planet is held to 0. The authors' rationale appeals to a short (on the order of tens of Myr) circularization timescale, the age of the star ($\sim$2 Gyr), a low stellar rotation rate ($\sim$0.03 d), and observational constraints favoring small eccentricities. If we allow for even a small eccentricity in our models, however, we find that we do not need to appeal to \textit{ad hoc} removal of iron at specific longitudes \citep[the favored explanation of][]{wardenier2021wasp76} to produce the observed blueshift signature. Because WASP-76b subtends more than $30\degr$ of phase during its transit, including a small eccentricity can both increase the magnitude of our blueshift at every point in transit and produce a trend consistent with the \cite{ehrenreich2020nightside} results.

Using the \texttt{exoplanet} package \citep{exoplanet} to model planet-frame RV deviations from a circular orbit within a $\chi^2$ fitting routine (see Appendix \ref{appendix:eccentricity-fitting} for fitting details), we find that the data are best explained by an eccentricity of \addressresponse{$e = 0.018$} and a longitude of periastron \addressresponse{$\omega=125\degr$} for a cloudless atmosphere.\footnote{As noted by \cite{showman2012doppler}, the quantity $e\cos(\omega)$, which can be derived from transit or secondary eclipse observations, is more constraining of an anomalous center-of-transit blueshift than $e$ alone.}

But eccentricity alone cannot produce the data's characteristic jump in Doppler shift post-ingress that our optically thick cloud models reproduced. Combining 10 scale height $\rm Al_2O_3$ clouds and condensation with a small eccentricity produces this work's best qualitative model fit (Figure \ref{fig:ecc_phase}, with $e = 0.017$ and $\omega=117\degr$). The quantitative fit also improves correspondingly: When including eccentricity, the $\chi^2$ value for the clear case decreases from 401 to 121, and the $\chi^2$ value for the cloudy case decreases from 329 to 112. The blueshift at the correct phases is increased by the eccentricity trend, thereby fully reproducing both the magnitude and time dependence of the \cite{ehrenreich2020nightside} data.

\addressresponse{Both our cloudless and cloudy eccentricity-inclusive modeling results are consistent within 1$\sigma$ with the $\omega$ derived from \cite{fu2020hubble} ($62_{-82}^{+67}\degr$).} Furthermore, both estimates are well within the 3$\sigma$ limit derived by \cite{west2016three}.

While the above eccentricity argument is perhaps satisfying, it may not accurately represent reality. Increasing blueshift as a function of phase has also been observed in the ultra-hot Jupiter WASP-121b \citep{bourrier2020hot, borsa2021atmospheric}. It is perhaps more likely that there exists an underlying physical mechanism common to both WASP-121b and WASP-76b than their orbital configurations happening to be similar\addressresponse{; the parameter estimation of WASP-121b at discovery constrains $e<0.07$ at 3$\sigma$ \citep{delrez2016wasp121}, and later studies have held the planet to a circular orbit \citep{evans2016detection,evans2017ultrahot,mikal2019emission,borsa2021atmospheric}. Furthermore, \cite{bourrier2020hot} place a tighter constraint on eccentricity ($e<0.0078$) at 3$\sigma$ from TESS data. It is yet to be determined whether a post-processed GCM could account for the blueshift trend in WASP-121b without modifications such as added eccentricity (e.g., as in the weak drag, deep RCB case in Fig.~\ref{fig:phase_results_iron_only}).} Hence, the question of whether cloudless models or the addition of optically thick clouds alone could provide a population-level explanation of phase-resolved phenomena warrants further investigation. 

\subsubsection{Non-hydrostatic effects}\label{sec:hydrostatic}
As explored in Section~\ref{sec:compare_tabernero}, the presence of non-hydrostatic effects in WASP-76b's atmosphere is supported by the behavior of the Ca II H\&K lines observed in \cite{tabernero2021espresso}. In turn, it is possible that a hydrodynamic outflow may be responsible for some portion of the anomalous blueshift signature.

By virtue of its close proximity to its host star, WASP-76b should experience strong UV radiation flux; such irradiation can result in mass loss in its upper layers \citep[e.g.,][]{erkaev2007roche, des2010evaporation, linsky2010observations, ehrenreich2015giant}. Indeed, \cite{seidel2021wasp} recently found evidence for high-altitude vertical winds in WASP-76b, which may deliver material to a hydrodynamic outflow (or serve as the base of the outflow itself) in the planet's exosphere. As proposed for WASP-121b \citep{bourrier2020hot}, an anisotropic expansion of the atmosphere could lead to variable Doppler shifts over the course of transit. 

Both our ray-striking radiative transfer and our GCM presuppose that WASP-76b's atmosphere is in local hydrostatic equilibrium. While our assumption may be valid strictly in the pressure regimes that we consider in our study --- simulations of atmospheric loss generally begin in the nanobar regime \citep[e.g.,][]{murrayclay2009escape, bourrier2015radiative, bourrier2016evaporating}, whereas the upper boundary of our atmosphere is at a pressure of $\approx$ 11~$\mu$bar --- enforcing the hydrostatic assumption prevents us from exploring this hypothesis. Fully non-hydrostatic models that include winds driven by UV radiation would be required to determine whether a physically plausible outflow could become optically thick enough at the correct phases to meaningfully contribute to the WASP-76b Doppler shift trend. 

\subsection{Limitations of our model \label{sec:model_limitations}}

As with any modeling study, it is challenging to incorporate all of the relevant physical and chemical processes in a self-consistent manner.
It is worth briefly noting here several of the limitations of our modeling effort that may impact our ability to fully capture the physical state of WASP-76b's atmosphere.  

Firstly, our GCM employs a double-gray radiative transfer scheme. While this scheme has benefits compared to more accurate, nongray approaches (e.g., speed, ability to diagnose underlying dynamics), the heating and cooling predicted by double-gray radiative transfer is inherently limited in spectral regions of strong wavelength dependence, given that these types of models use two opacities as opposed to precise, wavelength-dependent opacities \citep{showman2009atmospheric, kataria2015atmospheric, lee2021simulating}. For example, such a modeling approach is limited in its ability to generate the strong thermal inversions predicted by 1-D and 3-D modeling of ultra-hot Jupiter atmospheres \citep[e.g.,][]{parmentier2018thermal, lothringer2019influence}. Additionally, while our post-processing scheme includes treatments of condensation and cloud formation, our GCM is cloud-free. This introduces a fundamental inconsistency between the two components of our modeling effort and misses the potential for radiative feedback from clouds, which can significantly impact the thermal structures of hot Jupiter atmospheres \citep[e.g.][]{lee2016dynamic,lines2018simulating,lines2019overcast,roman2019modeled, roman2021clouds}.

Secondly, as noted in our companion paper (\linktocite{maykomacek2021spitzer}{May \& Komacek et~al.} \citeyear{maykomacek2021spitzer}), one explanation for the inability of our models to fully reproduce the \cite{ehrenreich2020nightside} data in a self-consistent manner could be our GCM's approximate treatment of drag. If the regions probed by high-resolution spectroscopy contain atmospheric flow that is sufficiently coupled to the magnetic field \citep[i.e., there is enough ionization for plasma effects to become notable;][]{perna2010magnetic}, collisions between electron, ion, and neutral populations can produce drag effects that depart from our Rayleigh approximation. In particular, the strength of the drag should depend strongly on the local temperature (via the amount of thermal ionization), the direction of the flow relative the magnetic field \citep{rauscher2013three}, and the geometry of the global magnetic field \citep{batygin2014non}, and
in some cases these interactions can actually serve to accelerate the flow of neutrals \citep{koskinen2014electrodynamics,rogers2017constraints,hindle2021observational}. \addressresponse{Recently, \cite{beltz2021exploring} found that accounting for locally calculated magnetic drag in a double-gray GCM of WASP-76b significantly altered the dayside atmospheric flow, driving flow toward the poles at low pressures and strongly reducing the extent of the equatorial jet.} The remaining question --- whether magnetic acceleration could account for the gap between our models and the \cite{ehrenreich2020nightside} data --- is an excellent avenue for future work, as ultra-hot Jupiters have strong atmospheric thermal ionization and hence high electrical conductivity, and if they have significant dipolar magnetic fields, they will almost certainly experience strong Lorentz forces.

Finally, the accuracy of our models at the spectrum level is bounded by our treatment of chemistry: Both the GCM and the chemical equilibrium models computed with \texttt{FastChem} assume solar metallicity and the absence of any disequilibrium processes. These assumptions may impact the thermal structure of our GCM; enhanced (i.e., greater than solar) metallicity in GCMs has been shown to promote greater day-night temperature contrasts \citep[e.g.,][]{showman2009atmospheric, kataria2015atmospheric} --- which could in turn drive day-night flow and affect our blueshift conclusions or alter chemistry \citep{steinrueck2019effect, drummond2020implications}. Furthermore, our condensation treatment does not account for a ``cold trap'' scenario \citep[e.g.,][]{showman2009atmospheric, spiegel2009can}. Specifically, our temperature-pressure profiles (Figure~\ref{fig: t-p profs}) intersect some condensation curves at multiple pressures; our approach to modeling condensation can in some cases remove gas-phase abundance at a deep pressure but allow it at a higher pressure. This is not physically plausible, as a cold trap would make the lower-pressure regions thermodynamically inaccessible to a gas-phase species if higher-pressure regions condense it out. 

From an observational perspective, \cite{fu2020hubble} indicate that the emission spectrum of WASP-76b at low resolution is consistent with solar metallicity and equilibrium chemistry. This result motivates our not accounting for disequilibrium, although the upper atmosphere on WASP-76b's dayside could experience significant photochemistry by virtue of its intense stellar insolation \citep[e.g.,][]{line2010high, moses2011disequilibrium, lavvas2014electron, molaverdikhani2019cold, shulyak2020stellar}. Thus, while our modeling approach does not explore the entire relevant chemical parameter space, it does not run contrary to previous conclusions about the state of WASP-76b's atmosphere.

\section{Conclusion}\label{conclusion}

To examine the hypothesis of iron rain on WASP-76b and delve into the planet's 3-D thermal, chemical, and wind structures, we post-process a grid of GCMs with a ray-striking radiative transfer code at high spectral resolution. We further apply a variety of post-processing schemes to account for condensation and/or rainout of iron and other species. Our major findings are as follows:

\begin{enumerate}
    \item Reproducing the large blueshifts observed by \citet{ehrenreich2020nightside}, \citet{kesseli2021confirmation}, and \citet{tabernero2021espresso} in the context of our GCMs almost certainly requires that WASP-76b's atmosphere is in a weak-drag state and has a deep RCB (i.e., a low internal temperature). These two conditions are necessary to drive the strong wind speeds and significant east-west limb asymmetries that are required to produce large blueshifts ($\gtrsim$~5~km\,s$^{-1}$) during transit.  
    \item With our cloud-free models, we are unable to reproduce the magnitude and time dependence of the anomalous iron blueshift signature reported in \cite{ehrenreich2020nightside} in a way that is consistent with the thermal and wind structure predicted by our GCM output. This conclusion is robust even when accounting for gas-phase iron condensation. We are therefore unable to reproduce the behavior of the \citet{ehrenreich2020nightside} toy model when applying our self-consistent GCM-based modeling approach.
    \item The maximum (egress) magnitude of the observed blueshift \textit{can} be reproduced if we allow for a small and previously unaccounted-for eccentricity of \addressresponse{$e \approx 0.017$}, which is within the bounds allowed by previous observational constraints \citep{fu2020hubble}.
    \item Including an optically thick cloud up to 10 scale heights thick in our post-processing scheme allows us to reproduce the sharp jump in Doppler shift prior to mid-transit that is evident in the \cite{ehrenreich2020nightside} and \cite{kesseli2021confirmation} data. While the cloud composition does not strongly affect the Doppler signature, (gray) $\rm Al_2O_3$ clouds provide the best fit\addressresponse{, with Fe clouds and $\rm Mg_2SiO_4$ clouds fitting marginally worse}. Similarly, while including gas-phase condensation along with thick clouds provides the best fit, the fit improvement by doing so is marginal. Combining these clouds with a small eccentricity accurately reproduces both the magnitude and the time-dependent trend of the \cite{ehrenreich2020nightside} Doppler shift observations. Notably, the cloud does not actually reach 10 scale heights at most locations in the atmosphere, as the atmospheric temperature-pressure structure can become too hot for our modeled clouds at high altitudes.
    \item Our model spectra line heights and Doppler shifts are broadly consistent with the \cite{tabernero2021espresso} analysis of the WASP-76b ESPRESSO data for a variety of chemical species. We find that individual line Doppler shifts can differentiate between drag timescale scenarios, whereas the altitudes at which these lines form cannot. Phase-resolved, cross-correlation-derived Doppler shifts are uniquely suited for differentiating between drag timescale scenarios and condensation/cloud scenarios, as they probe many lines (and hence altitudes) over multiple geometries.
    \item Ca II is distinct among the studied chemical species. Its line heights as observed by \cite{tabernero2021espresso} are likely explained by non-hydrostatic effects (i.e., atmospheric escape).  Its modeled Doppler shift is distinctly redder than all other investigated species, and its phase-resolved Doppler shift seems to be strongly dependent on the assumed planetary temperature structure.
\end{enumerate}

Overall, we note broad qualitative agreement between our GCM outputs (specifically those with weak drag and deep RCBs) and the existing high-resolution data for WASP-76b, as presented by \citet{ehrenreich2020nightside}, \citet{kesseli2021confirmation}, and \citet{tabernero2021espresso}. However, quantitatively reproducing both the observed magnitude and time dependence of the Doppler shift data is more elusive. Specifically,
our models call into question that the large blueshifts of iron can be simply explained by condensation of this species out of the gas phase. Our work instead indicates that WASP-76b's anomalous blueshift during transit cannot solely be interpreted as iron condensation, insofar as the thermal and wind structures output by our GCM accurately predict the physical state of the planet. Instead, we have invoked a combination of condensation, optically thick clouds, and a small but non-zero eccentricity to best explain the \citet{ehrenreich2020nightside} data in particular. Other aspects of the WASP-76b high-resolution data \citep[i.e., the results presented by][]{tabernero2021espresso} are well-explained by a larger subset of our forward models. We point out that our challenges in reproducing the observed Doppler shift signatures from \citet{ehrenreich2020nightside} do not call into question the fidelity of the data themselves, which have been independently verified across two instruments, three studies, and six nights of observation. 

WASP-76b remains a well-studied and interesting ultra-hot Jupiter, and we have shown in this work that the high-resolution data are highly diagnostic of 3-D processes in the planet's atmosphere. Ultimately, additional data for this planet and for other ultra-hot Jupiters with similar properties, along with improved 3-D and 1-D modeling approaches, will shed light on the relative importance of clouds, condensation, and atmospheric circulation in shaping the properties of this intriguing class of exoplanet.

\begin{acknowledgements}

A.B.S., E.M.-R.K., and E.R. acknowledge funding from the Heising-Simons Foundation. J.L.B. acknowledges support from NASA XRP grant 80NSSC19K0293. M.M. (Mansfield) acknowledges funding from the NASA FINESST program.

The authors acknowledge the University of Maryland supercomputing resources (\url{http://hpcc.umd.edu}) made available for conducting the research reported in this paper.

This research made use of \texttt{exoplanet} \citep{exoplanet} and its dependencies \citep{exoplanet:arviz, exoplanet:astropy13, exoplanet:astropy18, exoplanet:pymc3, exoplanet:theano}.

This research has made use of NASA’s Astrophysics Data System Bibliographic Services.

We thank Rodrigo Luger and Dan Foreman-Mackey for their insightful Theano-related suggestions. We also thank Prof. David Ehrenreich for making available the stellar-frame Doppler velocities used for Figures \ref{fig:phase_results_iron_only}, \ref{fig:all_species_ccf}, and \ref{fig:ecc_phase}. We further thank Joost Wardenier for providing limiting-case spectra of WASP-76b for comparison.

\addressresponse{Finally, we thank the anonymous reviewer for their helpful comments that greatly improved the quality of this manuscript.}
\end{acknowledgements}

\software{\texttt{astropy} (\citealt{astropy:2018}), \texttt{batman} \citep{kreidberg2015batman}, \texttt{Corner} \citep{corner}, \texttt{emcee} \citep{foreman2013emcee}, \texttt{exoplanet} \citep{exoplanet}, \texttt{FastChem} \citep{stock2018fastchem}, \texttt{IPython} (\citealt{perez2007ipython}), \texttt{Matplotlib} (\citealt{hunter2007matplotlib}), \texttt{NumPy} (\citealt{2020NumPy-Array}), \texttt{Numba} \citep{lam2015numba},
\texttt{pandas} (\citealt{mckinney2010data}),  
\texttt{SciPy} (\citealt{virtanen2020scipy}), \texttt{tqdm} (\citealt{da2019tqdm})}

\appendix

\section{Eccentricity fitting} \label{appendix:eccentricity-fitting}
Our procedure to fit our eccentricity-inclusive model to the \cite{ehrenreich2020nightside} blueshift data is as follows:

\begin{enumerate}
    \item Initialize an \texttt{exoplanet} \texttt{KeplerianOrbit} object with stellar and orbital parameters from \cite{ehrenreich2020nightside}. Additionally, compile a \texttt{Theano} \citep{exoplanet:theano} function associated with the \texttt{KeplerianOrbit} object that takes eccentricity and longitude of periastron as inputs.
    \item Write a likelihood function for our two orbital parameters (eccentricity and longitude of periastron) and a parameter characterizing the degree of error bar underestimation (log(f)), assuming Gaussian errors on the data.
    \item Identify the global minimum of the negative log likelihood using the \texttt{minimize} routine from \texttt{SciPy}. We find a minimum at \addressresponse{$e=0.019$, $\omega=123.1\degr$, log(f) = $-1.84$} for our cloudless model and a minimum \addressresponse{at $e=0.018$, $\omega=115.6\degr$, log(f) = $-1.89$} for our optically thick cloud model.
    \item Initialize 32 \texttt{emcee} walkers in a small Gaussian ball around the global minimum identified in the previous step. Repeat this step separate runs with two different priors: a uniform prior over the relevant parameter ranges and a prior using the constraints provided by \cite{fu2020hubble} (see Table \ref{table:mcmc}).
    \item Run the \texttt{emcee} walkers for 20,000 steps.
    \item Visually identify whether the walkers effectively explore the parameter space via trace plots.
    \item Check that the sampler's acceptance fraction is between 10\% and 90\%. All of our chains have acceptance fractions near 62\%.
    \item Discard the first few multiples of the autocorrelation time as ``burn-in.'' We generally discard 1,000 samples, which is more than sufficient for this criterion.
    \item Thin the samples by a factor of half the autocorrelation time ($\approx 20$) to generate reasonably independent samples for the posterior distribution.
    \item Assess convergence using the \cite{geweke1992evaluating} criterion.
\end{enumerate}

\begin{figure}
    \centering
    \includegraphics[scale=0.6]{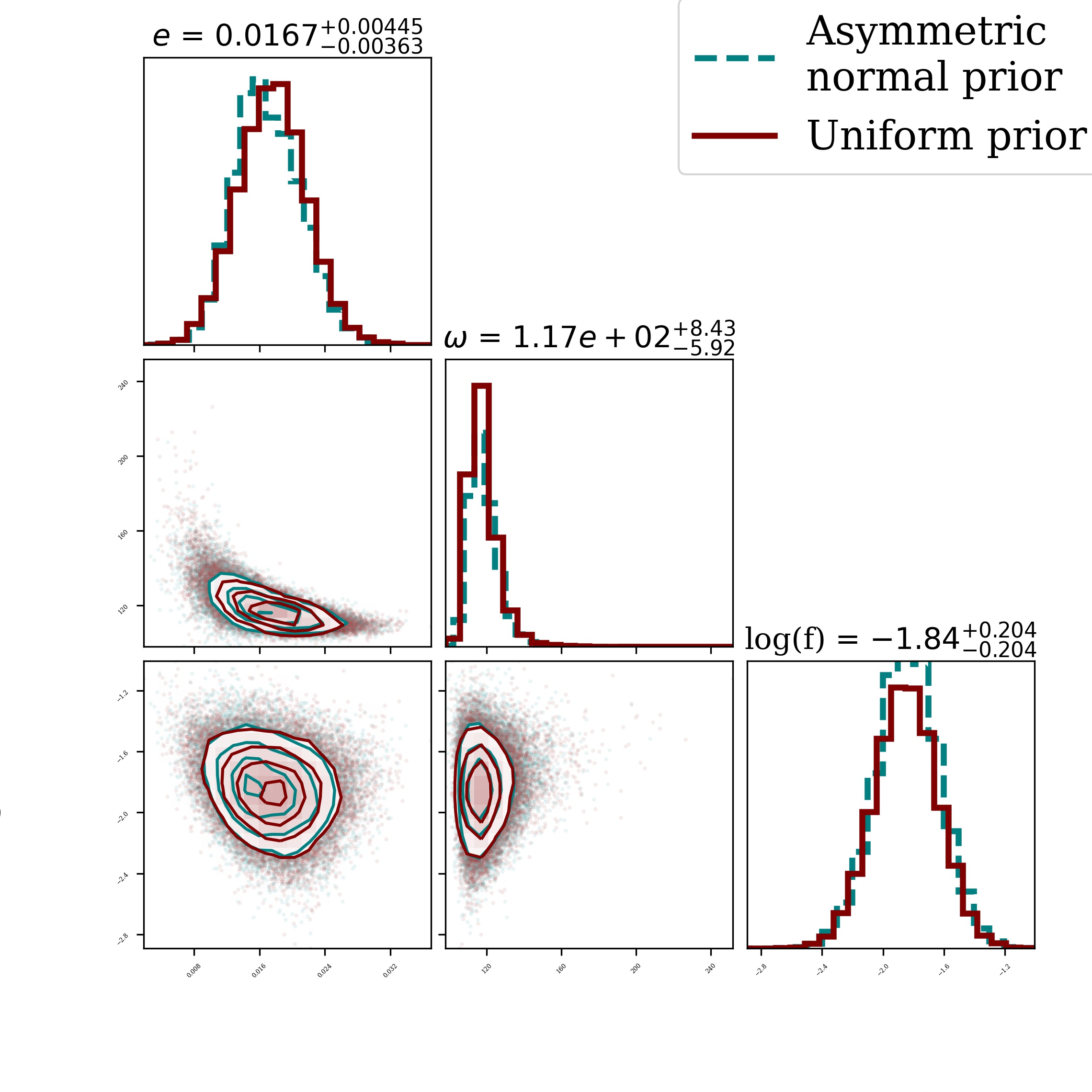}
    \caption{Posterior corner plot corresponding to our 10 scale height $\rm Al_2O_3$ cloud MCMC run. Both priors result in similar posteriors, implying that the posterior is heavily influenced by the data. Quantities are reported for the asymmetric normal prior (drawn from the \citealt{fu2020hubble} constraints).}
    \label{fig:ecc_posteriors}
\end{figure}

We find that our results are relatively insensitive to our choice of prior (of the two explored; Figure~\ref{fig:ecc_posteriors}) and pass our diagnostic tests.

As a separate note, both our cloudless and cloudy MCMC runs converge on similar median eccentricities. The eccentricities in both runs meet the $e > 2.45\sigma_e$ metric for statistically significant nonzero eccentricity defined by \cite{lucy1971spectroscopic}, which accounts for there being zero phase to explore at exactly zero eccentricity.

\begin{table}
\caption{MCMC values}
\label{table:mcmc}
\centering
\begin{tabular}{cccccc}
\toprule
\textbf{Model} &
\textbf{Param} & \textbf{Priors} &\textbf{Fit value}&\textbf{Autocorr. length}&\textbf{Indep. samples}
    \\

\bottomrule
   \makecell{\renewcommand\cellalign{ccccccc}No clouds, \\weak drag, \\deep RCB}&
    $e$ & \makecell{\renewcommand\cellalign{ccccccc}$\unif(0,1)$,\\$\mathcal{P}$(0.016, 0.011, 0.018)\footnote{The scri   pt $\mathcal{P}$(a,b,c) indicates Gaussian priors of standard deviations b (lower) and c (upper) joined at their median, a.}}& \addressresponse{$0.018^{+0.004}_{-0.004}$} & \addressresponse{46, 46} & \addressresponse{433, 434} \\
    --- & 
    $\omega$ &\makecell{\renewcommand\cellalign{cc}$\unif(0\degr,180\degr)$,\\$\mathcal{P}$(62, 82, 67)}& \addressresponse{$125^{+9.87}_{-6.87}$\degr} & \addressresponse{47, 48} & \addressresponse{417, 409}\\
    ---&
    log(f) & $\unif(-10,1)$ &\addressresponse{$-1.79^{+0.20}_{-0.20}$} & \addressresponse{41, 43}& \addressresponse{485, 459} \\\hline
    \makecell{\renewcommand\cellalign{ccccccc}$\rm Al_2O_3$ clouds, \\weak drag, \\deep RCB,\\cond. on}&
    $e$ & \makecell{\renewcommand\cellalign{ccccccc}$\unif(0,1)$,\\$\mathcal{P}$(0.016, 0.011, 0.018)}& \addressresponse{$0.017^{+0.004}_{-0.004}$} & \addressresponse{44, 46} & \addressresponse{450, 432}\\
    --- & 
    $\omega$ &\makecell{\renewcommand\cellalign{cc}$\unif(0\degr,180\degr)$,\\$\mathcal{P}$(62, 82, 67)}& \addressresponse{$117^{+8.43}_{-5.92}$\degr} & 48, 43 & 412, 457 \\
    ---&
    log(f) & $\unif(-10,1)$ & \addressresponse{$-1.84^{+0.20}_{-0.20}$} & \addressresponse{39, 42} & \addressresponse{504, 467}\\\hline

\end{tabular}
\end{table}


\bibliography{sample63}{}
\bibliographystyle{aasjournal}


\end{document}